\documentclass[letterpaper,journal]{IEEEtran}
\usepackage{amsmath,amsfonts}
\usepackage{amsmath}
\usepackage{algorithm}
\usepackage{algpseudocode}
\usepackage{array}
\usepackage[caption=false,font=normalsize,labelfont=sf,textfont=sf]{subfig}
\usepackage{textcomp}
\usepackage{stfloats}
\usepackage{url}
\usepackage{verbatim}
\usepackage{graphicx}
\usepackage{cite}
\usepackage{algorithm}
\usepackage{algpseudocode}
\usepackage{graphicx}
\usepackage{multirow}
\usepackage{amsmath}
\usepackage{bm}
\usepackage{amsfonts}
\usepackage{url}
\usepackage{comment}
\usepackage{tabularx}
\usepackage{framed}
\usepackage{array}
\usepackage{longtable}
\usepackage{rotating}
\usepackage{booktabs}
\usepackage{indentfirst}
\usepackage{fancyhdr}
\usepackage{framed}
\usepackage{colortbl}
\usepackage{tcolorbox}
\usepackage{tabularx}
\usepackage{makecell}
\usepackage{diagbox}
\usepackage{multicol} % 导言区添加
\usepackage{courier} % 加载 Courier 字体作为等宽字体
\usepackage{enumitem} % 引入 enumitem 包

\begin{document}

% \title{Obfuscation Makes Learning Difficult: A Robust Framework for Malware Family Classification}

\title{When Label Noise Meets Class Imbalance: A Robust Framework for Android Malware Family Classification}

\author{Haolan Zhang , Cuiying Gao , Fulin Zhao , Heng Li , Haoran Wang , Chang Luo , Tiejun Wu, Hui Shu*, Wei Yuan* 
        % <-this % stops a space
\thanks{This paper was produced by the IEEE Publication Technology Group. They are in Piscataway, NJ.}% <-this % stops a space
\thanks{Haolan Zhang, Cuiying Gao, Fulin Zhao, Heng Li, Haoran Wang, Chang Luo and Wei Yuan are with the School of Electronic Information and Communications, Huazhong University of Science and Technology, Wuhan 430074, China}
\thanks{Cuiying Gao is also with JD.com, Beijing 100000, China.}
\thanks{Heng Li is also with The Hong Kong Polytechnic University, Hong Kong 999077, China.}

\thanks{Tiejun Wu is with the NSFOCUS Technologies Group Co., Ltd., Beijing 100000, China}
\thanks{Hui Shu is with Key Laboratory of Cyberspace Security, Ministry of Education, Zhengzhou, 450000, China}
\thanks{*Corresponding author: Wei Yuan (e-mail: yuanwei@mail.hust.edu.cn), Hui Shu (e-mail: shuhui163@126.com)}%
\thanks{Manuscript received April 19, 2021; revised August 16, 2021.}}

\markboth{Journal of \LaTeX\ Class Files,~Vol.~14, No.~8, August~2021}%
{Shell \MakeLowercase{\textit{et al.}}: A Sample Article Using IEEEtran.cls for IEEE Journals}

\IEEEpubid{0000--0000/00\$00.00~\copyright~2021 IEEE}

\maketitle

\begin{abstract}
Machine learning methods for Android malware family classification have achieved high accuracy, but their application is hindered by two major challenges. First, the widely used code obfuscation severely disrupts the automated labeling process and introduces substantial label noise into training datasets. Second, training data sets often exhibit severe class imbalance, leading to poor performance of family classification models. 
Although existing studies have proposed various solutions to either label noise or class imbalance, they often overlook the interplay between these two factors. Under class imbalance, the presence of hard-to-learn minority-class samples can significantly impair the effectiveness of existing countermeasures for noisy samples. To jointly address label noise and class imbalance, we propose a robust Android malware family classification framework \texttt{RoMaC}. It employs a self-training strategy to correct noisy labels and, more importantly, discriminately treats head-family and tail-family samples. This design effectively mitigates the adverse impact of class imbalance on noise-robust learning. Moreover, \texttt{RoMaC} integrates a class reweighting mechanism with multi-model ensemble learning, thereby enhancing both classification accuracy and noise robustness.
We evaluate \texttt{RoMaC} on a combined dataset of two public datasets.
When 30\% samples are obfuscated, \texttt{RoMaC} achieves an overall Macro-F1 score of 0.803, accuracy of 0.871, and tail-class Macro-F1 score of 0.672, accuracy of 0.784. Compared to existing methods, \texttt{RoMaC} demonstrates 6\%–20\% performance improvements across various obfuscation scenarios and noise levels.
\end{abstract}


\section{INTRODUCTION}\label{SEC:INTRO}
As the most widely used mobile operating system, Android faces increasingly severe malware threats due to its openness and wide usage \cite{MalwareFox2025}. Android malware family classification, which identifies the behavioral patterns and characteristics of similar malware variants to group them into the same family, has become an effective approach to analyzing and mitigating malware threats. Existing classification methods predominantly employ supervised learning, achieving classification based on shared features within malware families \cite{mahdavifar2022effective,zhang2023detecting,islam2023android}. Their performance heavily depends on high-quality datasets. In practice, however, the wide use of code obfuscation and the intrinsic distribution characteristics of malware often result in datasets plagued by \textit{label noise} and \textit{class imbalance}. 

\begin{figure}[t]
  \centering
  \includegraphics[height=0.15\textheight]{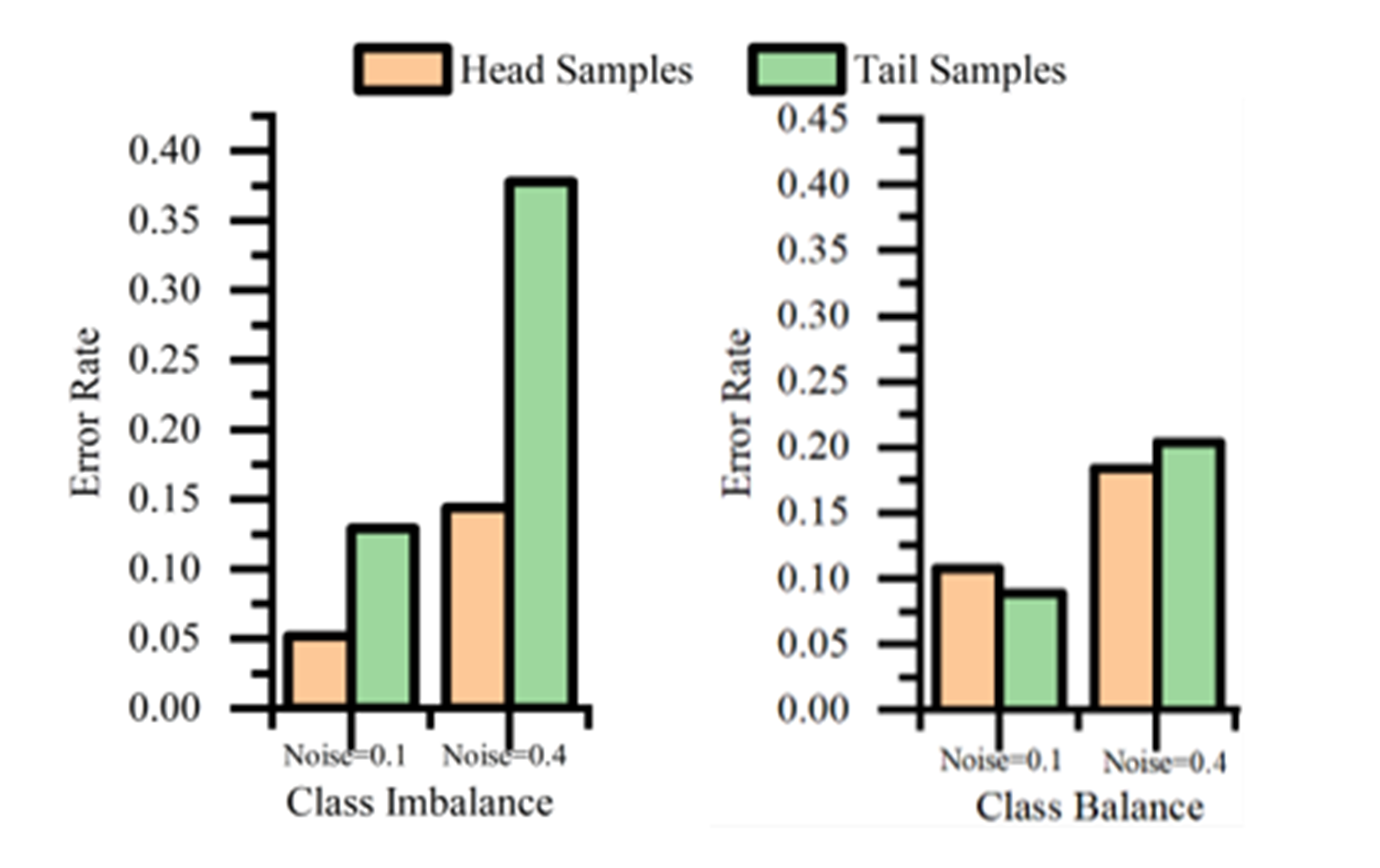} % 调整宽度并指定文件名
  \caption{Pseudo-label error rates in balanced and imbalanced scenarios across two noise levels.}
  \label{dongji}
\end{figure}
Code obfuscation is a technique that deliberately alters code structure or behavioral characteristics to hinder reverse engineering and feature extraction while preserving program functionality \cite{cuiying2024uncovering}\cite{dong2018understanding}.
It can be abused to evade Android malware detection \cite{gao2023obfuscation} through modifying application bytecode. To counteract this, robust features resistant to obfuscation, such as those based on sensitive function calls or Dalvik opcode sequences \cite{gao2023obfuscation}, have been developed for malware detection. However, malware detection methods also heavily rely on high-quality labels of training samples. At present, Android application labeling primarily relies on automated platforms (e.g., VirusTotal) to generate AV labels \cite{sebastian2016avclass} and derive family labels accordingly, which are susceptible  to code obfuscation. Unfortunately, over 60\% of malware samples are obfuscated, making the family labels of 60\%–75\% of malware incorrect \cite{cuiying2024uncovering}\cite{dong2018understanding}.

\IEEEpubidadjcol
Class imbalance also complicates malware family classification. This phenomenon indicates a few head families (e.g., Android banking trojans or adware families) account for the majority of samples while numerous tail families (typically small-scale or regional malware) have very few samples \cite{sawadogo2022android,almomani2021android}. For instance, in CICMalDroid \cite{mahdavifar2020dynamic}, the dominant or head family ‘fakeinst’ includes 1,530 samples, whereas the tail family ‘hiddenapp’ contains only 14 samples. Class imbalance makes model learning biased, overfitting to head classes (families) and poorly generalizing to tail classes, degrading malware family classification performance \cite{liu2021imbalance}.

Most existing countermeasures to label noise and class imbalance have two main drawbacks. First, they address either label noise or class imbalance in isolation, making them often fail under complex and realistic conditions \cite{liu2021imbalance, qi2024family, canbek2022gaining}. This is because the fundamental approach to mitigating label noise relies on model self-training to generate pseudo-labels for further training, while class imbalance disrupts this process, causing the model to overly favor head classes and generate incorrect pseudo-labels.
% On the one hand, label noise is commonly corrected through model self-training, where pseudo-labels generated with high model confidence are used for further training. Yet, class imbalance disrupts this mechanism by causing the model to favor head classes, resulting in incorrect pseudo-labels for tail-class samples. 

As shown in Fig. \ref{dongji}, our experiments show that the pseudo-label error rate for tail families is increased significantly by class imbalance under different label noise levels.
Second, addressing class imbalance usually involves assigning higher weights to tail-class samples, which heavily depends on weight selection. However, the presence of label noise distorts the distribution of malware families and enhances the difficulty in weight selection, leading to increased weight tuning overhead and degraded classification performance.

% traditional reweighting methods attempt to address class imbalance by assigning higher weights to samples from the tail class. However, their effectiveness is highly dependent on accurate weight estimation. Label noise complicates this process, leading to degraded model performance and increased tuning overhead.

In this paper, we propose an Android malware family classification framework \texttt{RoMaC} to mitigate the effects of obfuscation-induced label noise and class imbalance, through investigating the interplay between label noise and class imbalance. The design philosophy of \texttt{RoMaC} is briefly summarized as follows. First, to address the issue of rising pseudo-label error rates caused by class imbalance, we propose a new semi-supervised self-training algorithm, which incorporates consistency regularization while assigning lower weights to uncertain pseudo-labels, effectively suppressing error propagation. 
% This approach significantly improves classification accuracy and robustness under class imbalance. 
Second, to tackle the challenge of precisely setting weights for head-class and tail-class samples under label noise, we introduce a model ensemble approach where each model exhibits distinct preferences toward tail classes and head classes and independently identifies malware families.
This preference is reflected through distinct weighting strategies for head classes and tail classes. By fusing model predictions via majority voting, \texttt{RoMaC} captures complementary discriminative cues learned under varying degrees of attention to tail classes. Our experiments confirm that this method markedly improves malware family classification performance under various label noise levels. 
% The underlying principle is that noisy samples affect individual models, but the multi-model ensemble operation effectively dilutes such impact, thereby enhancing noise robustness. Moreover, this method reduces the risk 
%  and tuning effort of determining optimal weights. 
The core idea is that under a diverse reweighting mechanism, the prediction errors of each model are differently affected by header class bias and labeling noise. As a result, incorrect predictions are usually inconsistent across models and easily diluted by the integration mechanism, while correct predictions are consistent and can be accumulated over time, thus improving robustness to class imbalance and label noise.

% The \texttt{RoMaC} framework introduces an innovative architecture that effectively addresses the intertwined challenges of label noise and class imbalance, offering a robust and efficient solution for complex malware classification scenarios. 

The architecture of \texttt{RoMaC} comprises four main modules: feature extraction, sample diversion, label correction, and ensemble reweighting. The feature extraction module adopts the existing feature extraction methods used for malware family classification. This explains why \texttt{RoMaC} is a general framework and does not rely on specific detection features.
In the sample diversion and label correction modules, our semi-supervised self-training algorithm is used to tackle label noise under class imbalance. In the ensemble reweighting module, various classes (i.e, malware families) are assigned different weights and a model ensemble is conducted to address class imbalance under label noise.

The main contributions of this work are summarized below. 

\begin{itemize}[leftmargin=0.3cm, topsep=0pt, partopsep=0pt] % 调整左边距
    \item \textit{Insight}. For the first time, we study the coupling and interplay of label noise and class imbalance in Android malware family classification. Our study demonstrates that label noise and class imbalance should be considered simultaneously rather than in isolation, which reveals a new direction for future research on malware family classification.
    \item \textit{Method}. We propose an Android malware family classification framework \texttt{RoMaC} to simultaneously counter label noise and class imbalance. Its key ideas include: 1) mitigating the negative impact of class imbalance during pseudo-label generation (to resist label noise), and 2) leveraging ensembles of diverse classification models to overcome the challenge of precise class weighting (to counter class imbalance).
    \item \textit{Performance}. For evaluation, extensive experiments are conducted on a combined dataset of two public datasets with varying levels of obfuscation. Experimental results demonstrate the superiority of \texttt{RoMaC} over existing approaches. Notably, under the most challenging setting with a 70\% obfuscation noise ratio, \texttt{RoMaC}  achieves a macro-F1 score of 0.528 and an accuracy of 0.721, surpassing state-of-the-art methods by 6\%–20\%.
    
    % \item We identify and systematically study the entangled challenge of label noise and class imbalance, and empirically reveal the significant negative impact of class imbalance on self-training noise-robust learning methods. This observation provides new insights and a foundation for further research on effective learning under such complex supervision conditions.

    % \item We investigate the core limitations of class reweighting strategies and propose a novel ensemble-based solution. This approach substantially improves the overall robustness to class imbalance, while more importantly, it eliminates the high risk and tuning burden associated with relying on a single fixed weighting scheme.
\end{itemize}

\section{BACKGROUND}
Android malware family classification utilizes APK features, such as code structure and permission requirements, to identify families like trojans and adware \cite{wu2019malscan,arp2014drebin,MaMaDroid}. It reveals malware attack patterns, and contributes to defense design and attack tracing. Currently,  machine learning models have been widely used for malware family classification \cite{han2020android,rashid2025hybrid,sahu2023deep}. In realistic environments, Android malware family classification needs to overcome the following two challenges. 
\subsection{Class Imbalance}
\begin{figure}[htbp]
  \centering
  \includegraphics[height=0.15\textheight]{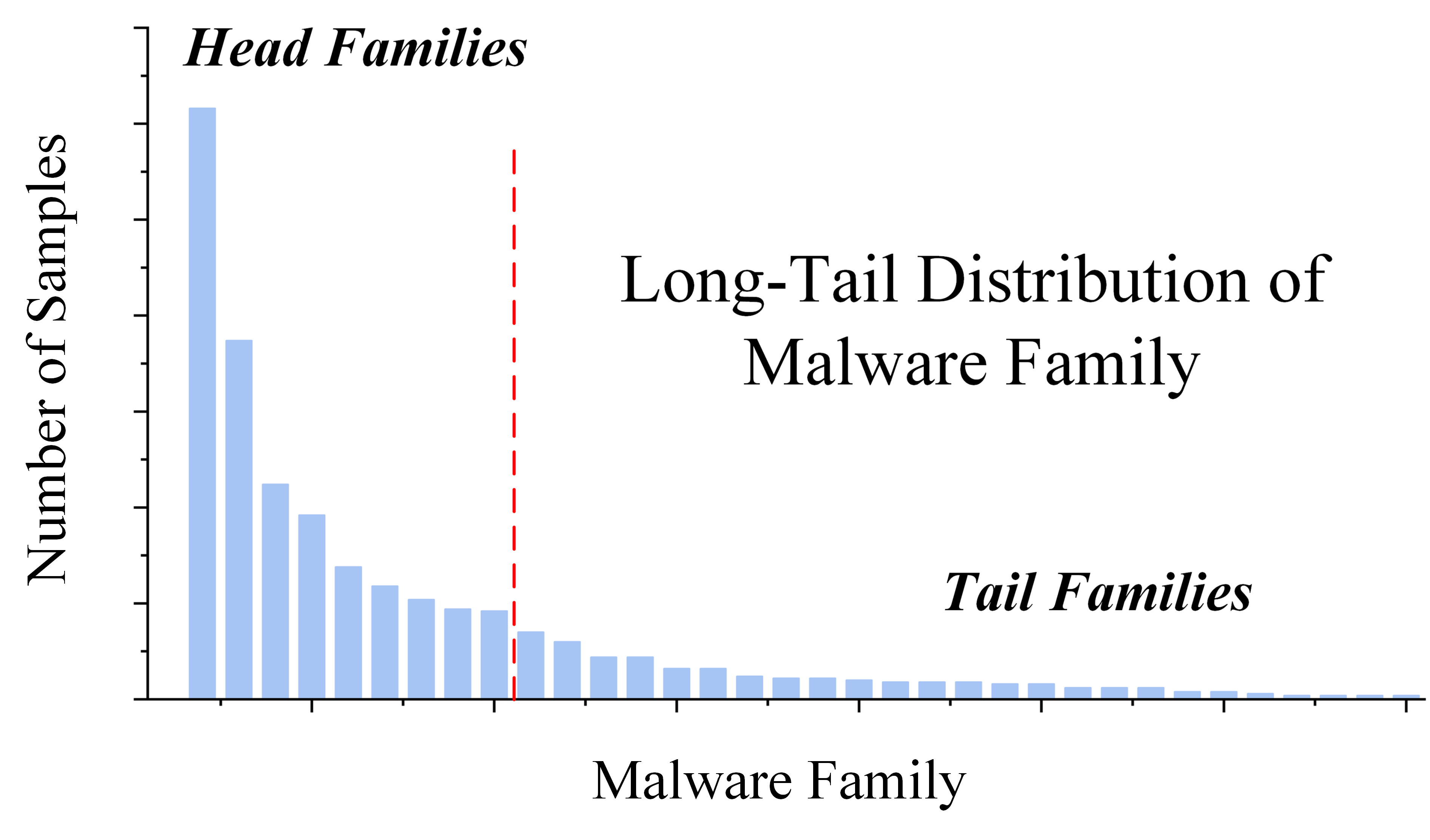} % 调整宽度并指定文件名
  \caption{Long-tail distribution of malware family.}
  \label{longatil}
\end{figure}

% Class imbalance markedly characterizes the distribution of malware families with a long-tail pattern. 
Class imbalance is frequently encountered in Android malware family classification. As illustrated in Fig. \ref{longatil}, the distribution exhibits a long-tailed pattern, where a small number of head classes dominate, while the majority of tail classes are severely underrepresented, often with only 1–2 samples \cite{mahdavifar2020dynamic,mahdavifar2022effective,zhou2012dissecting}. This imbalance arises from the dynamics of the Android malware ecosystem: head classes, often developed by sophisticated cybercrime organizations with extensive dissemination networks, dominate the sample pool, whereas tail families — typically produced by individual developers with limited propagation capabilities — remain scarce \cite{sawadogo2022android}. 

Previous work \cite{liu2024learning} shows that such imbalance biases the model toward head-class samples, thereby reducing recall for tail-class samples. This weakens the model's generalizability and may lead to overestimated accuracy during evaluation \cite{liu2021imbalance,bai2020unsuccessful}. In particular, tail-class samples often correspond to high-risk or emerging threats, and failing to detect them can have serious consequences.

\subsection{Code Obfuscation}\label{SUBSEC:CB}
% Code obfuscation is a technique that modifies code structure or behavioral characteristics to hinder reverse engineering and feature extraction while preserving program functionality . 
Common code obfuscation methods include CSE (Constant String Encryption), CID (Call Indirection), and CR (Class Rename) \cite{li2018significant,zhou2012dissecting}. Malware developers can take advantage of specialized tools such as Obfuscapk \cite{aonzo2020obfuscapk} to change features such as API calls or behavioral patterns, thus avoiding malware detection. Dong et al. \cite{dong2018understanding} analyze 114,560 Android apps and find that approximately 63.5\% of malicious apps in real-world datasets employ obfuscation techniques, highlighting the strong preference for obfuscation of malware developers.

Current malware family labeling relies heavily on AV labels \cite{sebastian2016avclass} generated by scanning platforms such as VirusTotal. However, code obfuscation introduces systematic perturbations into APK features, severely impairing the reliability of automated labeling. For example, CSE renders feature-based detection engines ineffective by encrypting key strings; CR disrupts semantic detection by altering symbol names. As a result, detection engines frequently misclassify obfuscated samples, severely contaminating the sample labeling process. It has been shown that different obfuscation techniques can alter the assigned malware family labels for 60\% to 75\% of samples \cite{gao2023obfuscation,cuiying2024uncovering}, introducing substantial label noise into training datasets. Therefore, it is imperative to handle code obfuscation in Android malware family classification.

%% 太重复了
% Researchers must collect and label malware samples to build a training dataset to deploy the system. However, the training dataset faces dual quality challenges. First, malware authors extensively employ code obfuscation techniques, resulting in mislabeling of obfuscated samples by automated tagging. Second, the scarcity of samples from high-threat malware families causes severe class imbalance in the training dataset. Specifically, the high cost of manual labeling necessitates reliance on automated labeling, but code obfuscation disrupts feature extraction and matching, leading to label noise. Furthermore, the limited availability of high-threat family samples, due to collection difficulties, results in severe class imbalance in the dataset, further exacerbating the learning bias of classification models.

% Our goal is to develop a general Android malware family classification framework for real-world scenarios. We attempt to tackle two critical challenges in the training data: class imbalance and significant label noise caused by code obfuscation. Second, our system improves the detection of high-threat, small-scale malware families, effectively reducing real-world security risks.

\section{METHODOLOGY}
In this section, we develop an Android malware family classification framework robust against label noise and class imbalance. Formally, let $(x_i,y_i)$ be the pair of an Android malware sample $x_i$ and its true label $y_i$ (which means the family of the malware). The inputs of malware family classification are a training data set $\widetilde{D}_{\text {train }}=\left\{\left(x_i, \widetilde{y}_i\right)\right\}_{i=1}^N$ and a testing data set $D_{t e s t}=\left\{x^{t e s t}\right\}$, where $\widetilde{y}_i$ is a noisy label that may be inconsistent with $y_i$. The sizes of different families may vary by tens or even hundreds of times. Our goal is to accurately infer the label of $x^{test}$ by using $\widetilde{D}_{\text {train }}$.

\subsection{Motivation}
\begin{figure*}[htbp]
  \centering
  \includegraphics[width=0.95\textwidth,height=0.25\textheight]{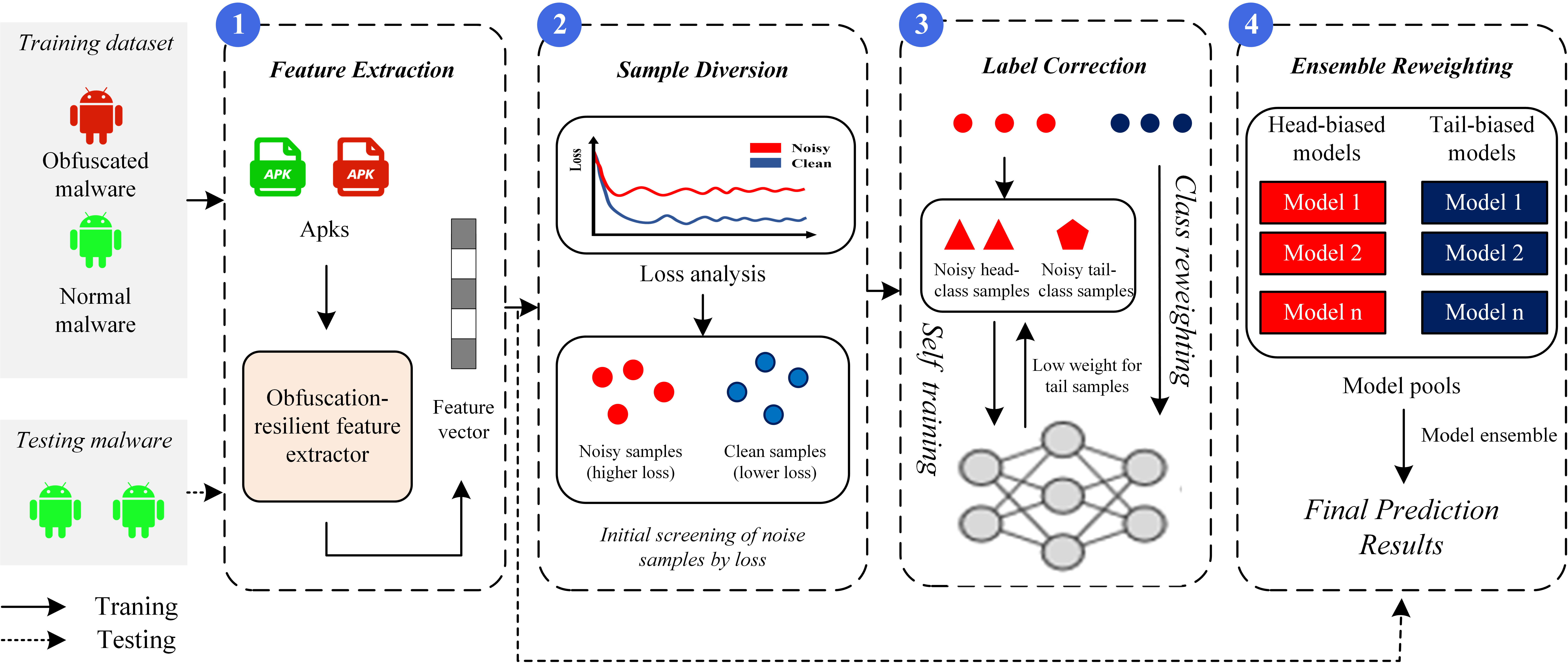} % 调整宽度并指定文件名
  \caption{The overview of \texttt{RoMaC}, which consists of four modules: Feature Extraction, Sample Division, Lable Correction and Ensemble Reweighting.}
  \label{fig:overview}
\end{figure*}

% 如前所述，类不平衡会导致为抵抗标签噪声而设计的自训练方法失效，特别是针对尾部类样本，自训练方法会生成大量错误伪标签。另一方面，标签噪声的存在进一步扭曲不同家族的分布，使类加权等传统类不平衡解决方案难以实施或难以取得令人满意的性能。

As mentioned earlier, class imbalance can lead to the failure of self-training methods designed to resist label noise, particularly for tail-class samples where self-training tends to generate numerous incorrect pseudo-labels. On the other hand, the presence of label noise further distorts the distributions of malware families, making traditional class imbalance solutions (e.g., class re-weighting) difficult to implement or hard to achieve satisfactory performance.

% The deep coupling of label noise and class imbalance poses severe challenges to traditional algorithms: class imbalance causes conventional self-training noise correction methods to fail, particularly by generating numerous erroneous pseudo-labels for minority tail-class samples; meanwhile, label noise further distorts data distributions, making classic imbalance solutions such as re-weighting difficult to achieve a globally optimal balance.

%为了克服上述挑战，该框架率先引入基于预训练损失的样本分流机制，主动剥离可疑样本标签以构建净化数据池，并通过自训练方法迭代地为未标记样本分配伪标签。

To address the first challenge, \texttt{RoMaC} employs a pre-training loss-based sample splitting mechanism, which proactively isolates the samples with suspicious labels to construct a purified data pool, while iteratively assigning pseudo-labels to unlabeled samples via self-training. To improve the accuracy of pseudo-labels, \texttt{RoMaC} applies differentiated learning strategies to unlabeled head-class and tail-class samples. More specifically, it conducts weak regularization on unlabeled tail-class samples with extremely low weights to effectively block the propagation of incorrect pseudo-labels; labeled samples are re-weighted to correct distributional bias.

%它将多个分类模型集成起来再对家族分类结果进行投票，其中每个模型对于class reweighting的策略是不同的，有的更偏向于头部类，而有的更偏向于尾部类。虽然我们始终无法知道最优的weights是什么，但模型集成的好处在于：当多个模型分类正确时，它们会给出一致的分类结果；而当多个模型分类错误时（由于weights不合理），它们往往会给出不同的分类结果。正确的decision由于票数较多将会脱颖而出而称为最终分类结果.相反，错误的decisions在fusion时会因为票数少而被忽略。

To overcome the second challenge, \texttt{RoMaC} proposes a novel ensemble reweighting method, which integrates multiple classification models and then votes on the classification results. Note that each model employs different strategies for class reweighting—some favoring head classes while others prioritize tail classes. Although it is hard for every model to optimally set the weights for various classes, the advantage of model ensemble lies in: when multiple models classify correctly, they yield a consistent decision; whereas when multiple models err (due to different weights), they often produce different decisions. When the decisions of all models are fused, the correct decision will stand out due to majority votes and become the final classification result. Contrarily, those erroneous decisions are ignored during voting because their corresponding votes are few. Therefore, the ensemble reweighting method strengthens the correction classification result and weakens the erroneous decisions, ultimately improving classification accuracy.
% Hence, this method significantly enhancing robustness and classification performance under class imbalance and label noise. 
% In the following subsections, we will detail these methods adopted in \texttt{RoMaC}.

\subsection{Architecture}
\texttt{RoMaC} consists of four main modules, including a feature extraction module, a sample diversion module, a label correction module (HeadTailMatch), and an ensemble reweighting module, as shown in Fig. \ref{fig:overview}. 

\textbf{Feature Extraction}: \texttt{RoMaC} extracts common static features from Android APK files such as OMM \cite{gao2023obfuscation}, Drebin \cite{arp2014drebin}. These features have been widely used in Android malware analysis. It should be pointed out that \texttt{RoMaC} is highly flexible and not limited to specific features.

\textbf{Sample Diversion}: \texttt{RoMaC} divides samples into two categories (i.e. labeled and unlabeled), based on their loss values computed after a pre-training phase. Specifically, \texttt{RoMaC} adopts the Exponential Moving Average (EMA) for training loss as a dynamic and robust criterion. Samples with a higher EMA loss are treated as unlabeled, while others are considered labeled.

\textbf{HeadTailMatch}: In \texttt{RoMaC}, we propose a novel self-training algorithm HeadTailMatch. For unlabeled samples, we design different optimization strategies for head-class and tail-class samples. For labeled samples, we introduce a class reweighting method based on family distribution to mitigate the problem of insufficient sample representation of rare but high-threat families.

\textbf{Ensemble Reweighting}: In general, it is challenging for a single-class reweighting strategy to simultaneously achieve high accuracy for rare families and maintain generalizability for common families. To address this, \texttt{RoMaC} introduces model ensemble and fuses the predictions of head-biased models and tail-biased models, hence mitigating the effects of class imbalance under label noise. 

\subsection{Feature Extraction and Sample Diversion} 
% \texttt{RoMaC} is designed with a core objective of enhancing and remaining compatible with existing robust feature representations, such as OMM, which analyzes Smali opcode sequences and constructs Markov transition matrices to effectively resist feature perturbations introduced by code obfuscation.
% Crucially, the core innovation and performance advantage of 

\texttt{RoMaC} does not redefine new features but serves as a general framework capable of seamlessly absorbing various robust features. 
When various features are used in the feature extraction module, \texttt{RoMaC} consistently demonstrates superior robustness and higher accuracy compared to baseline methods. Therefore, feature extraction is not the focus of our method and will not be elaborated here. Extensive experiments presented later in this paper validate the generality of the feature extraction module in \texttt{RoMaC}. 

% Regardless of the feature extraction module employed,  RoMaC consistently collaborates with these features. It demonstrates sustained improvements in robustness and classification accuracy over baseline methods, even under highly obfuscated conditions. These results affirm RoMaC's high compatibility with different feature representations and its broad utility as a universal optimization layer for robust malware classification.

Now we consider the second module, that is, the sample diversion module. It is used to filter out noisy labels in the training dataset. The key challenge lies in designing a robust criterion to effectively distinguish clean samples from noisy samples. 
% This differentiation aims to maximize the use of clean data while minimizing the influence of noisy labels during training. 
% However, a significant limitation arises when training solely on the identified clean labeled data: the exclusion of the entire data pair $(x, y)$ associated with an incorrect label $y$ leads to the forfeiture of potentially valuable insights regarding the underlying data distribution $p(x)$. Therefore, it is crucial to leverage noisy data. We propose treating samples with noisy labels as unlabeled data, allowing their reintegration into the training process. 
The loss associated with clean samples is typically lower than that of noisy samples \cite{han2018co}. This discrepancy arises because clean labels provide consistency during gradient updates, enabling more effective model fitting. Consequently, instantaneous loss has been widely used to differentiate between noisy and clean samples \cite{shen2019learning}. However, the instantaneous loss is an unstable signal, subject to rapid fluctuations due to the inherent randomness in model training, hence often leading to cumulative errors throughout the training process. 

To address this, we use the Exponential Moving Average (EMA) loss as a more stable criterion for identifying noisy samples. For a sample $(x_i, y_i)$, we characterize its loss dynamics using a simple exponential moving average of the instantaneous loss $\ell\left(f\left(x_i; \theta_t\right), y_i\right)$, where $f\left(x_i; \theta_t\right)$ represents the model output, and $\theta_t$ is the model parameters at training step $t$. The loss is defined and computed recursively as 
\begin{equation}
l_{t+1}(i)= \gamma \times \ell\left(f\left(x_i ; \theta_t\right), y_i\right)+(1-\gamma) \times l_t(i) 
\end{equation}
where $\gamma \in[0,1]$ is a discounting factor. We update the EMA loss solely for the samples diversion, utilizing the byproduct $l_t(i)$ of training without necessitating additional inference.

\begin{figure}[htbp]
  \centering
  \includegraphics[width=0.45\textwidth,height=0.15\textheight]{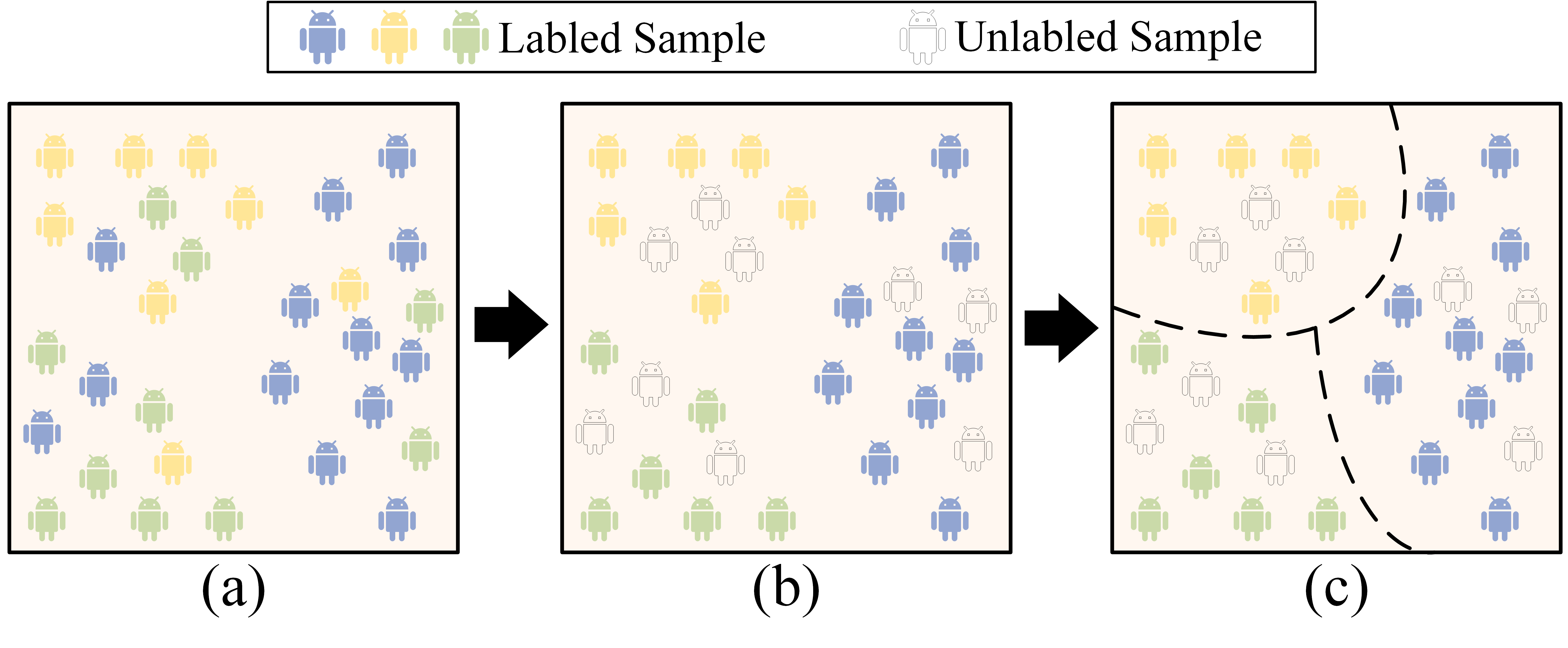} % 调整宽度并指定文件名
  \caption{Sample diversion and HeadTailMatch: (a) Noisy data. (b) Transformed data. (c) HeadTailMatch.}
  \label{SSL}
\end{figure}

We adopt the cross-entropy function as the loss function for our model. After pretraining, we identify the top $d$ samples with the highest EMA loss, designating them as uncertain samples and subsequently removing their labels. The remaining samples are treated as labeled samples. As shown in Fig. \ref{SSL}, sample diversion categorizes the samples into labeled and unlabeled samples. Note that we cannot completely exclude unlabeled samples from model training, since they may provide potentially valuable insights regarding the underlying sample distribution. In \texttt{RoMaC}, we design a self-training algorithm HeadTailMatch, which is applied to the transformed data to correct erroneous labels by assigning pseudo-labels to unlabeled samples.  In the next subsection, we will discuss HeadTailMatch in detail. 

% in order to maximize data utilization and enhance the model's robustness against label noise.

% The pre-trained model retains its parameters and proceeds to the next training phase.

\subsection{Label Correction}

% Considering the intertwined effects of label noise and class imbalance,  HeadTailMatch counters the adverse effects introduced by erroneous labels while maintaining robustness against severe class imbalance. 
HeadTailMatch builds on the FixMatch \cite{sohn2020fixmatch} framework, which combines pseudo-labeling and consistency regularization for semi-supervised learning. Unlike FixMatch, HeadTailMatch is tailored for imbalanced and noisy data by applying class reweighting and assigning different strategies for head-class and tail-class unlabeled samples.
The main workflow of HeadTailMatch is detailed in Algorithm 1. 
The final loss $\mathcal{L}$ consists of three components: the unsupervised loss for head classes $\mathcal{L}_h$, the unsupervised loss for tail classes $\mathcal{L}_t$, and the supervised loss $\mathcal{L}_x$.

\begin{algorithm}
\caption{HeadTailMatch}
\begin{algorithmic}[1]
\Require Labeled dataset $\mathcal{X}$, unlabeled dataset $\mathcal{U}$, number of total training epochs $K$, confidence threshold $\tau$, weak regularization weight $\lambda_{weak}$, learning rate $\alpha$, reweighting coefficient $\mu$, pre-training model $P$.
\Ensure Well-trained model $f(\cdot;\Theta)$.
\State Initialize the weights $\Theta$ for the model $f(\cdot)$ by $P$.
\For{$k = 1$ to $K$}
    \State Partition $\mathcal{U}$ into unlabeled head dataset $\mathcal{H}$, and unlabeled tail dataset $\mathcal{T}$.
    \State Sample a random batch $\left\{x_b, y_b\right\}_{b=1}^{B_x} \sim \mathcal{X}$, $\left\{u_b^h\right\}_{b=1}^{B_h} \sim \mathcal{H}$, $\left\{u_b^t\right\}_{b=1}^{B_t} \sim \mathcal{T}$.
    \For{$b = 1$ to $B_h$}
        \State // Get the weakly augmented version $h_w$ and strongly augmented version $h_s$.
        \State $q = f(h_w; \Theta)$.  // The prediction probability vector
        \State $\widehat{q} = \arg\max(q)$. // The confidence of pseudo-labels
        \State $\mathcal{L}_h = \frac{1}{B_h} \sum_{b=1}^{B_h} H(\widehat{q}, f(h_s)) I(\max(q) \geq \tau)$.  // Unsupervised head loss
    \EndFor
    \For{$b = 1$ to $B_t$}
        \State // Get the weakly augmented version $t_w$ and strongly augmented version $t_s$.
        \State $p = f(t_w; \Theta)$. 
        \State $\widehat{p} = \arg\max(p)$.
        \State $\mathcal{L}_t = \lambda_{weak}\frac{1}{B_t} \sum_{b=1}^{B_t} H(\widehat{p}, f(t_s)) I(\max(p) \geq \tau)$. // Unsupervised tail loss
    \EndFor
    \For{$b = 1$ to $B_x$}
        \State $\omega_c=n_c^{-\mu}$. // The weight of $c_{th}$ class
        \State $\mathcal{L}_x = \frac{1}{B_x} \sum_{b=1}^{B_x} \omega_c H\left(y_b, f\left(x_b\right)\right)$. // Supervised loss
    \EndFor
    \State $\mathcal{L} = \mathcal{L}_h + \mathcal{L}_t + \mathcal{L}_x$. // Final loss function
    \State Update the model’s weights $\Theta$ by minimizing the loss function $\mathcal{L}$.
    \State // Optional: Decay the learning rate $\alpha$.
\EndFor
\end{algorithmic}
\end{algorithm}

By sample diversion, the input dataset $\widetilde{D}_{\text {train }}$ is transformed into a labeled dataset $\mathcal{X}$ and an unlabeled dataset $\mathcal{U}$. HeadTailMatch employs different strategies for them. For unlabeled data, HeadTailMatch employs consistency regularization and self-trained pseudo-labels to enable effective training. Samples \cite{sohn2020fixmatch,laine2016temporal} with different labels are separated in low-density regions and nearby samples are expected to produce consistent predictions. In particular, applying weak augmentations to unlabeled data should not significantly alter the model’s outputs. Based on this principle, we use the predictions from weakly augmented samples as pseudo-labels and compute the unsupervised regularization loss against the the predictions from strongly augmented data thus improving the generalization ability of the model model through consistency regularization.

Existing pseudo-labeling methods typically employ a confidence threshold to ensure pseudo-label quality \cite{sohn2020fixmatch,zhang2021flexmatch,han2018co}. Pseudo-labels are derived from the model’s self-training predictions, which may contain errors during early training stages or when feature distributions are complex \cite{chen2022debiased,liu2022acpl,wang2024towards}. Code obfuscation exacerbates feature distribution complexity and introduces label noise, degrading pseudo-label reliability. The confidence threshold mandates that only predictions with probabilities above a certain level are used as pseudo-labels. However, in the presence of class imbalance, a simple confidence threshold is inadequate for addressing the challenges of pseudo-label assignment. We observe that, even with high-confidence constraints, pseudo-labels for unlabeled tail samples exhibit a significant error rate. This problem arises because head classes, with their abundant samples, dominate the formation of the model’s decision boundaries, causing the model to erroneously assign tail samples to head classes with high confidence. Consequently, a simple confidence threshold fails to distinguish these erroneous high-confidence pseudo-labels. To address this, we use the weak regularization constraint of low weight for the unlabeled tail loss. 

Specifically, for the unlabeled head data $u_b^h$, we first apply data augmentation to generate a strongly augmented version $h_s$ and a weakly augmented version $h_w$, thus transforming the batch of unlabeled head data into a batch with pseudo-labels, i.e., ($u_b^h$,$\widehat{q}$). In HeadTailMatch, the pseudo-label is derived from the model’s prediction on the weakly augmented version, with $\widehat{q}$ representing the pseudo-label, specifically the index of the maximum value within the probability vector $q$ obtained from the model’s prediction. Mathematically, $\arg\max(q)$ can be interpreted as the confidence level of the pseudo-label. In addition, to ensure the quality of pseudo-labels, we set the confidence threshold $\tau$ = 0.95. If $\arg\max(q) < \tau$, the corresponding data and pseudo-label are excluded. Subsequently, for the retained high-confidence data, we compute the cross-entropy loss between the model’s output on the strongly augmented version and the pseudo-label, resulting in the loss for the unlabeled head data. This unsupervised loss for head classes is defined as
\begin{equation}
\mathcal{L}_h=\frac{1}{B_h} \sum_{b=1}^{B_h} H\left(\widehat{q}, f\left(h_s\right)) I(\max (q) \geq \tau)\right..
\end{equation}
here $f(h_s)$ represents the model’s output following strong augmentation of the sample. $H$ represents the cross-entropy function, $B_h$ indicates the number of samples per batch, and the function $I$ serves as a binary indicator function, producing a value of 1 when the condition within parentheses is satisfied. Through this function, HeadTailMatch ensures that an unlabeled head data sample $u_b$ is incorporated into the internal model parameter update process only if the model’s output ($q = f(h_w; \Theta)$) demonstrates high confidence.

For unlabeled tail samples, the high-confidence pseudo-labels may still be incorrect. Following the unsupervised loss formulation for head samples exacerbates these errors, amplifying model bias and reducing the detection capability for rare but high-threat families. To this end, based on the consistency regularization and pseudo-labeling techniques, we impose a weak regularization constraint of low weight $\lambda_{weak}$ on the tail samples, defining the tail loss to be
\begin{equation}
\mathcal{L}_t = \lambda_{weak} \frac{1}{B_t} \sum_{b=1}^{B_t} H\left(\widehat{p}, f(t_s)\right) I(\max(p) \geq \tau).
\end{equation}

To mitigate the risk of erroneous pseudo-labels from unlabeled tail samples complicating model training and degrading performance, we usually assign a small value to $\lambda_{weak}$ (See more details in Section IV D). 

For the labeled dataset $\mathcal{X}$, we mitigate the impact of imbalanced malware family distribution using a class reweighting method. The core idea is to assign different weights to each class based on the number of samples from the corresponding family, ensuring that the model gives more attention to tail families while avoiding domination by head families during training. Specifically, we assign higher weights to tail family samples and lower weights to head family samples, thereby balancing the contribution of each class to the supervised loss. The loss is defined as
\begin{equation}
\mathcal{L}_x = \frac{1}{B_x} \sum_{b=1}^{B_x} \omega_c H\left(y_b, f\left(x_b\right)\right).
\end{equation}
\begin{equation}
\omega_c=n_c^{-\mu}.
\label{jiaqun}
\end{equation}

Eq. \ref{jiaqun} is used for class reweighting, where $\omega_c$ represents the weight assigned to the samples of the $c$-th class, $n_i$ denotes the number of samples in the $c$-th class, and $\mu$ is a reweighting parameter that quantifies the extent of sample reweighting. As $\mu$ increases, the model progressively prioritizes tail samples, whereas smaller $\mu$ values shift the model’s focus toward head samples. To guide the training of our model, we combine the three losses to formulate the final loss.
\begin{equation}
\mathcal{L} = \mathcal{L}_h + \mathcal{L}_t + \mathcal{L}_x.
\end{equation}

\textbf{Data Augmentation}: To apply consistency regularization, data augmentation is required. Since image data augmentation techniques are unsuitable for malware data, we propose a general data augmentation method. Given a sample $\mathbf{x} \in \mathbb{R}^d$, we first generate a mask vector $\mathbf{m}=$ $\left[m_1, \ldots, m_d\right]^T \in \mathbb{R}^d$ where $m_d$ is randomly sampled from a Bernoulli distribution with probability $p$. Secondly, we compile a feature library using the complete training set data. We can then conduct data augmentation based on the following equation.
\begin{equation}
\tilde{\mathbf{x}}=\mathbf{x} \odot \mathbf{m}+(1-\mathbf{m}) \odot \overline{\mathbf{x}}.
\label{aug}
\end{equation}
where $\tilde{\mathbf{x}}$ denotes the augmented data and $\overline{\mathbf{x}}$ signifies synthetic data generated through random sampling across all dimensions within the feature library. For example, given a dataset with three samples [\texttt{1},\texttt{2},\texttt{3}], [\texttt{4},\texttt{5},\texttt{6}], and [\texttt{7},\texttt{8},\texttt{9}], the resulting augmented data becomes [random(\texttt{1},\texttt{4},\texttt{7}), random(\texttt{2},\texttt{5},\texttt{8}), random(\texttt{3},\texttt{6},\texttt{9})]. As is evident from Eq. \ref{aug}, our novel augmentation approach substitutes specific feature values with those of other samples, better reflecting the properties of the real-world sample. For example, each dimension in Malscan \cite{wu2019malscan} features is within $[0, 1]$. The method applies to a wide range of malware features. The rich feature library improves the representativeness of the synthetic data and allows the model to better capture the distributional properties of the samples, thus improving the robustness of the classification.

\subsection{Ensemble Reweighting}
HeadTailMatch proposed in the last subsection can effectively handle label noise under class imbalance. In this subsection, we consider how to counter class imbalance when label noise exists. A common countermeasure for class imbalance is class reweighting, which adjusts the sample weights of different classes (e.g., weights in the loss function) during training to make the model focus more on tail classes. However, relying on a single reweighting strategy may have inherent limitations. Insufficient weighting for tail classes usually fails to correct the model’s bias toward tail classes, while excessive weighting may distort the data distribution and lead to overfitting. Furthermore, the presence of label noise further complicates the setting of appropriate class weights, thereby exacerbating the problem. 

Instead of relying on a single reweighting configuration, we propose an ensemble reweighting method to break this fundamental bottleneck. It jointly trains a set of base models, each of which employs a different reweighting strategy. These base models span a spectrum from conservative to aggressive weighting schemes, thereby producing diverse feature spaces with varying emphases on tail classes. This not only substantially improves the overall robustness of the model against imbalance, but also eliminates the risk and manual burden associated with optimizing weight setting. Moreover, ensemble learning inherently enhances robustness against label noise by introducing model diversity and aggregating model predictions (or decisions). As mentioned in Section III. A, noisy samples affect base models in different ways, leading them to produce diverse misclassification decisions. The majority vote among base models can ignore misclassification decisions and highlight correct classification decisions.

% An inappropriate $\mu$ may cause the model to overemphasize minority classes (tail classes), thus leading to a tail-biased local optimum. To address this, we propose an ensemble learning approach--imbalance ensemble, that aggregates predictions from head-biased and tail-biased models, leveraging their complementary strengths to enhance overall performance. 

Specifically, by varying the weight of the tail class denoted by $\mu$ (see Eq.\ref{jiaqun}), we generate diverse models with different reweighting degrees: larger values $\mu$ attach more importance to tail classes. 
In particular, when $\mu$ is sufficiently large, the model becomes overly biased toward tail classes, prioritizing their learning at the cost of neglecting head classes, and we refer to such models as tail-biased models. In contrast, when $\mu$ is small, the model only suppresses the weights of the head classes to a limited extent, failing to balance the learning focus between the head and tail classes. As a result, it remains skewed toward head classes, which we term head-biased models.
We construct a model pool trained with different $\mu$ values and integrate the complementary predictions of head-biased and tail-biased models via majority voting, enhancing overall performance and mitigating the effects of label noise.

%%后面没有修改

We select $\mu_1, \mu_2,...,\mu_n$ based on the diversity of the model to create a set of models $M_1, M_2,...,M_n$, as detailed in Section \ref{RQ3}.
\begin{equation}
P\left(c \mid x\right)=\sum_{t=1}^T \frac{1}{T} \cdot p_{t}(x).
\label{voting}
\end{equation}
\begin{equation}
\hat{y}=\arg \max _{c} P\left(c \mid x\right).
\end{equation}

We adopt a model ensemble via soft voting, where each base model contributes to the final prediction. Let $T$ denote the number of base models. By default, uniform weights are used to ensure equal contribution and emphasize model complementarity over individual performance. The weights can also be adjusted for specific scenarios to better adapt to varying dataset characteristics.

In Eq. \ref{voting}, $P\left(c \mid x\right)$ represents the final predicted probability for class $c$, while $p_{t}(x)$ denotes the probability assigned by the $t$-th classifier that sample $x$ pertains to class $c$. Ultimately, we designate the class with the highest average probability $\hat{y}$ as the predicted result. Mathematically, $\hat{y}$ is the class among all categories with the highest $P\left(c \mid x\right)$. 

Furthermore, ensemble learning significantly increases robustness against label noise. By synthesizing predictions from multiple base models, ensemble learning effectively mitigates classification biases caused by noisy labels and, when combined with class reweighting and consistency regularization, significantly enhances the handling of both class imbalance and label noise challenges.

\section{EVALUATION}

We evaluate \texttt{RoMaC} on public datasets by answering the four Research Questions (RQs).

\textbf{RQ1: Effectiveness}. Can \texttt{RoMaC} mitigate the obfuscation-induced label noise and class imbalance?

\textbf{RQ2: Superiority}. Does \texttt{RoMaC} consistently achieve superior performance compared to state-of-the-art approaches in various scenarios?

\textbf{RQ3: Hyperparameter Analysis}.
How does the selection of key hyperparameters affect model performance?

\textbf{RQ4: Ablation Analysis}. What are the contributions of the individual components within \texttt{RoMaC}?

% \textbf{RQ4: Time Efficiency}: How does the time efficiency of \texttt{RoMaC} compare with that of existing methods?

\subsection{Experimental Setup} 
\subsubsection{Dataset} We use two public datasets for evaluation.
\begin{itemize}[leftmargin=0.3cm, topsep=0pt, partopsep=0pt] % 调整左边距
    \item \textbf{CICMalDroid} \cite{mahdavifar2020dynamic} consists of 8,407 Android malware samples. These samples were collected between December 2017 and December 2018 from multiple sources, including the VirusTotal service, Contagio security blog, AMD, MalDozer, and other datasets utilized in recent research studies.
    
    \item \textbf{MalRadar} \cite{wang2022malradar} is a growing and up-to-date Android malware dataset constructed using the most reliable method: collecting malware based on security experts' analysis reports. The dataset contains 4,534 Android malware samples released from 2014 to April 2021.

\end{itemize}

 To evaluate \texttt{RoMaC}, we combined the CICMalDroid and MalRadar datasets to create a comprehensive dataset simulating real-world scenarios. The malware family labels are derived using AVClass \cite{sebastian2016avclass}, which aggregates labels from multiple anti-virus engines. First, AVClass normalizes the input labels. Second, it removes non-specific terms such as `Trojan' or `Generic' and merges aliases of the same family. Finally, it counts the frequency of occurrence of candidate family labels based on a voting mechanism, requiring at least two engines to agree on the final label. Due to sample detection failures and other reasons, the final dataset contains 11,455 samples, encompassing 74 families with an imbalance ratio of 119:1, meaning that the largest family contains 119 times more samples than the smallest one.
 % (see details in appendix \ref{stastic}).

In our experiments, we use the widely-used obfuscation tool Obfuscapk \cite{aonzo2020obfuscapk} to induce label noise. As a black-box obfuscation solution specifically designed for Android applications, Obfuscapk enables multiple obfuscation transformations on APK packages without requiring access to the source code.
Using Obfuscapk, we implement three types of obfuscation: ConstStringEncryption (\textbf{CSE}), encrypt and store string constants in the code; CallIndirection (\textbf{CID}), disrupt the original function call relationships through indirect calls; ClassRename (\textbf{CR}), replace the class/method names with meaningless identifiers. Considering real-world scenarios where multiple obfuscations coexist, we also explore the mixed obfuscation \textbf{(MIX)}, which combines all three types to simulate complex scenarios.
We define the \textbf{obfuscation noise ratio} as the proportion of samples that are obfuscated. Recent studies \cite{gao2023obfuscation,cuiying2024uncovering} show that over 60\% of Android malware samples employ obfuscation, with 65\% to 75\% of these obfuscated samples introducing label noise. According to these studies, the obfuscation noise ratio varies from 0.3 to 0.7 in our experiments, while the actual label noise ratio is estimated to be approximately between 0.2 and 0.5.

% This setup allows for a systematic evaluation of \texttt{RoMaC}'s classification performance under different levels of obfuscation noise.

\subsubsection{Baselines} 
We conduct a thorough evaluation by comparing \texttt{RoMaC} with existing label noise learning methods (MentorMix \cite{jiang2020beyond}), class imbalance learning methods  (LDAM \cite{cao2019learning}), label noise learning methods under class imbalance (MORSE \cite{wu2023grim}), and a standard DNN baseline. 

\subsubsection{Implementation} Our experiments were conducted on a hardware platform with an NVIDIA RTX 3060 GPU, an Intel i5-10400 hexacore processor operating at 2.6 GHz, and 32 GB of system memory.
We list all the hyperparameters used by \texttt{RoMaC} in Table \ref{tab:para}. For the feature extractor module, we choose the OMM feature \cite{gao2023obfuscation} to evaluate \texttt{RoMaC} except RQ1. This is because OMM, as an instruction-level feature, characterizes app behavior via Dalvik opcode transition probabilities and is inherently resilient to obfuscation techniques such as Reflection and Renaming, making it appropriate for evaluating \texttt{RoMaC}.
% Table generated by Excel2LaTeX from sheet '参数'
\begin{table}[htbp]
  \centering
  \caption{The Hyperparameter Settings of \texttt{RoMaC}}
    \begin{tabular}{c|c|c}
    \toprule
    Para. & Value & Description \\
    \midrule
    $\lambda_{weak}$     & 0.1   & Weak regularization weight \\
    $n$     & 6     & Number of ensemble models \\
    $d$     & 0.1   & Proportion of unlabeled data \\
    $\tau$     & 0.95  & Confidence threshold for pseudo-labels \\
    $r$     & 0.4   &  Proportion of head families \\
    $\eta$  &0.001  &  Learning rata\\
    % $p_{strong}$     & 0.7   & Strong augmentation parameter \\
    % $p_{weak}$     & 0.05  & Weak augmentation parameter \\
    \bottomrule
    \end{tabular}%
  \label{tab:para}%
\end{table}%

Prior research \cite{ren2018learning} shows that at high learning rates, reweighting can amplify the interference of noisy samples, biasing the model towards common families or causing convergence instability. In malware family classification, this bias can hinder the detection of rare but high-threat families. In contrast, a lower learning rate allows the model to adjust incrementally to the effects of reweighting, starting from an initial representation, thereby mitigating potential oscillations or overfitting risks. This approach enhances class distribution balance and improves robustness in handling imbalanced data and high-noise environments. Consequently, we adopt a lower learning rate throughout the model training process.

For the proportion of head families $r$, we use the truncation point of a fitted power-law distribution to get the most appropriate value of $r$.  Long-tailed data distributions are common in various domains such as vision, text, and security. A widely adopted approach to characterize such distributions is fitting them to a power-law function, where the frequency of a class is inversely proportional to its rank. Specifically, given a ranked list of classes in descending order of frequency, the power-law form is defined as $P(k) \propto k^{-\alpha}$, where $k$  is the rank of a class, $\alpha$ is the power-law exponent. In practice, many real-world datasets follow a truncated power-law due to limited data or domain-specific constraints: 
$P(k) \propto k^{-\alpha}e^{- \lambda k}$, where $\lambda$ is the truncation parameter. Based on this, the head-tail boundary is often defined using the cutoff point $k_{cutoff}$ = $\frac{1}{\lambda}$, and the ratio $r=\frac{k_{cutoff}}{C}$ (with $C$ is the total number of classes), reflects the proportion of head classes. In our dataset, the proportion of head families $r$ is approximately 0.4.

\subsubsection{Metrics} To evaluate \texttt{RoMaC} under class imbalance, we consider both all classes and tail classes. Specifically, we employ the macro F1 score (M-F1) and accuracy (ACC) as evaluation metrics. For each class, we can calculate the number of true positive samples (TP),  the number of false positive samples (FP), and the number of false negative samples (FN). Then we have $\text{ACC} = \frac{\sum_{i=1}^C \text{TP}_i}{N}$, where $\text{TP}_i$ is the number of true positives for $i_{th}$ class, and $N$ is the total number of samples, $C$ denotes the total number of classes in the dataset. 
As for the F1 Score, it is the harmonic mean of precision and recall, with $\text{Precision}_i = \frac{\text{TP}_i}{\text{TP}_i + \text{FP}_i}$ and  $\text{Recall}_i = \frac{\text{TP}_i}{\text{TP}_i+ \text{FN}_i}$. Macro-F1 is used to balance class performance, and can be calculated as $\text{Macro-F1} = \frac{1}{C} \sum_{i=1}^C 2 \cdot \frac{\text{Precision}_i \cdot \text{Recall}_i}{\text{Precision}_i + \text{Recall}_i}$. 

\subsection{RQ1:Effectiveness} 
\textbf{Goal and Setup}. Here we extract malware features using various methods and compare \texttt{RoMaC} with a baseline DNN model, validating \texttt{RoMaC}'s performance improvements under label noise and adaptability to various features. 
% The ob R means the obfuscation noise rate. 
The used features include three robust features (FARM \cite{han2020android}, OMM \cite{gao2023obfuscation}, and APIGraph \cite{zhang2020enhancing}) and two normal features (Drebin \cite{arp2014drebin} and Malscan \cite{wu2019malscan}).
Experiments are conducted under two scenarios: no obfuscation and high obfuscation-induced label noise rate (denoted by ob R ), in order to clearly demonstrate the practical impact of obfuscation-induced label noise on model training.

\textbf{Analysis and Results}. Table \ref{tab:WithDNN} illustrates the effectiveness of \texttt{RoMaC} under different obfuscation rates with various features. Using the no-obfuscation scenario as a baseline, relying solely on anti-obfuscation features fails to train high-performing models. Anti-obfuscation features (e.g., OMM and FARM) outperform conventional ones (e.g., Drebin and MalScan), as they partially resist the effects of feature distortion caused by code obfuscation. Nevertheless, when the obfuscation rate reaches 70\%, performance deteriorates notably across both feature types. For example, OMM experiences a 21.6\% drop in overall M-F1 (from 0.658 to 0.442), and a 13.8\% drop for tail classes. This significant degradation highlights the difficulty of maintaining reliable classification under high levels of obfuscation-induced label noise.

\texttt{RoMaC} demonstrates substantial performance improvements for all and tail classes. For instance, with OMM features, at the 70\% obfuscation ratio, \texttt{RoMaC} improves overall Macro-F1 and ACC by 8.6\% and 10.6\%, respectively, and tail-class Macro-F1 and ACC by 11.0\% and 7.2\%, respectively. Moreover, \texttt{RoMaC} effectively integrates multiple features, consistently achieving superior results, significantly addressing the challenges of obfuscation-induced label noise, and enhancing tail-class accuracy to alleviate class imbalance.

In addition, we also evaluate the effectiveness of \texttt{RoMaC} on a clean dataset (no obfuscation). Even on clean but class-imbalanced data, \texttt{RoMaC} achieves notable performance gains, with tail-class performance improving by approximately 10\%–15\%. This underscores \texttt{RoMaC}'s effectiveness on clean yet imbalanced datasets, demonstrating its robustness and superiority in real-world scenarios, regardless of obfuscation ratios.

% Table generated by Excel2LaTeX from sheet 'RQ1不同抗混淆特征'
\begin{table}[htbp]
  \centering
        \tabcolsep = 0.08cm % 修改列间距
  \caption{Evaluation of \texttt{RoMaC} on scenarios with 70\% obfuscation and no obfuscation using various features}
    \begin{tabular}{lcccccccc}
    
    \toprule
    \multicolumn{1}{c|}{\multirow{3}[6]{*}{System}} & \multicolumn{4}{c}{70\% obfuscation } & \multicolumn{4}{c}{No obfuscation} \\
\cmidrule{2-9}    \multicolumn{1}{c|}{} & \multicolumn{2}{c}{Full classes} & \multicolumn{2}{c}{Tail classes} & \multicolumn{2}{c}{Full classes} & \multicolumn{2}{c}{Tail classes} \\
\cmidrule{2-9}    \multicolumn{1}{c|}{} & M-F1  & ACC   & M-F1  & ACC   & M-F1  & ACC   & M-F1  & ACC \\
    \midrule
    \rowcolor[rgb]{ .867,  .922,  .969} FARM & \cellcolor[rgb]{ 1,  1,  1}0.451 & \cellcolor[rgb]{ 1,  1,  1}0.604 & \cellcolor[rgb]{ 1,  1,  1}0.386 & \cellcolor[rgb]{ 1,  1,  1}0.519 & \cellcolor[rgb]{ 1,  1,  1}0.645 & \cellcolor[rgb]{ 1,  1,  1}0.782 & \cellcolor[rgb]{ 1,  1,  1}0.517 & \cellcolor[rgb]{ 1,  1,  1}0.659 \\
    \rowcolor[rgb]{ .867,  .922,  .969} FARM+RoMaC & \cellcolor[rgb]{ 1,  1,  1}0.568 & \cellcolor[rgb]{ 1,  1,  1}0.708 & \cellcolor[rgb]{ 1,  1,  1}0.507 & \cellcolor[rgb]{ 1,  1,  1}0.596 & \cellcolor[rgb]{ 1,  1,  1}0.792 & \cellcolor[rgb]{ 1,  1,  1}0.869 & \cellcolor[rgb]{ 1,  1,  1}0.681 & \cellcolor[rgb]{ 1,  1,  1}0.812 \\
    \rowcolor[rgb]{ 1,  .949,  .8} OMM   & \cellcolor[rgb]{ 1,  1,  1}0.442 & \cellcolor[rgb]{ 1,  1,  1}0.615 & \cellcolor[rgb]{ 1,  1,  1}0.378 & \cellcolor[rgb]{ 1,  1,  1}0.521 & \cellcolor[rgb]{ 1,  1,  1}0.658 & \cellcolor[rgb]{ 1,  1,  1}0.783 & \cellcolor[rgb]{ 1,  1,  1}0.516 & \cellcolor[rgb]{ 1,  1,  1}0.658 \\
    \rowcolor[rgb]{ 1,  .949,  .8} OMM+RoMaC & \cellcolor[rgb]{ 1,  1,  1}0.528 & \cellcolor[rgb]{ 1,  1,  1}0.721 & \cellcolor[rgb]{ 1,  1,  1}0.488 & \cellcolor[rgb]{ 1,  1,  1}0.593 & \cellcolor[rgb]{ 1,  1,  1}0.794 & \cellcolor[rgb]{ 1,  1,  1}0.878 & \cellcolor[rgb]{ 1,  1,  1}0.682 & \cellcolor[rgb]{ 1,  1,  1}0.813 \\
    \rowcolor[rgb]{ .886,  .937,  .855} APIGraph & \cellcolor[rgb]{ 1,  1,  1}0.433 & \cellcolor[rgb]{ 1,  1,  1}0.587 & \cellcolor[rgb]{ 1,  1,  1}0.373 & \cellcolor[rgb]{ 1,  1,  1}0.484 & \cellcolor[rgb]{ 1,  1,  1}0.615 & \cellcolor[rgb]{ 1,  1,  1}0.743 & \cellcolor[rgb]{ 1,  1,  1}0.497 & \cellcolor[rgb]{ 1,  1,  1}0.618 \\
    \rowcolor[rgb]{ .886,  .937,  .855} APIGraph+RoMaC & \cellcolor[rgb]{ 1,  1,  1}0.532 & \cellcolor[rgb]{ 1,  1,  1}0.682 & \cellcolor[rgb]{ 1,  1,  1}0.461 & \cellcolor[rgb]{ 1,  1,  1}0.556 & \cellcolor[rgb]{ 1,  1,  1}0.769 & \cellcolor[rgb]{ 1,  1,  1}0.858 & \cellcolor[rgb]{ 1,  1,  1}0.649 & \cellcolor[rgb]{ 1,  1,  1}0.743 \\

    \rowcolor[rgb]{ .741,  .843,  .933} Drebin & \cellcolor[rgb]{ 1,  1,  1}0.428 & \cellcolor[rgb]{ 1,  1,  1}0.599 & \cellcolor[rgb]{ 1,  1,  1}0.348 & \cellcolor[rgb]{ 1,  1,  1}0.470 & \cellcolor[rgb]{ 1,  1,  1}0.610 & \cellcolor[rgb]{ 1,  1,  1}0.762 & \cellcolor[rgb]{ 1,  1,  1}0.482 & \cellcolor[rgb]{ 1,  1,  1}0.600 \\
    \rowcolor[rgb]{ .741,  .843,  .933} Drebin+RoMaC & \cellcolor[rgb]{ 1,  1,  1}0.522 & \cellcolor[rgb]{ 1,  1,  1}0.685 & \cellcolor[rgb]{ 1,  1,  1}0.465 & \cellcolor[rgb]{ 1,  1,  1}0.554 & \cellcolor[rgb]{ 1,  1,  1}0.772 & \cellcolor[rgb]{ 1,  1,  1}0.856 & \cellcolor[rgb]{ 1,  1,  1}0.663 & \cellcolor[rgb]{ 1,  1,  1}0.785 \\
    \rowcolor[rgb]{ 1,  .8,  1} Malscan & \cellcolor[rgb]{ 1,  1,  1}0.438 & \cellcolor[rgb]{ 1,  1,  1}0.598 & \cellcolor[rgb]{ 1,  1,  1}0.344 & \cellcolor[rgb]{ 1,  1,  1}0.484 & \cellcolor[rgb]{ 1,  1,  1}0.565 & \cellcolor[rgb]{ 1,  1,  1}0.758 & \cellcolor[rgb]{ 1,  1,  1}0.395 & \cellcolor[rgb]{ 1,  1,  1}0.579 \\
    \rowcolor[rgb]{ 1,  .8,  1} Malscan+RoMaC & \cellcolor[rgb]{ 1,  1,  1}0.516 & \cellcolor[rgb]{ 1,  1,  1}0.664 & \cellcolor[rgb]{ 1,  1,  1}0.424 & \cellcolor[rgb]{ 1,  1,  1}0.512 & \cellcolor[rgb]{ 1,  1,  1}0.667 & \cellcolor[rgb]{ 1,  1,  1}0.804 & \cellcolor[rgb]{ 1,  1,  1}0.625 & \cellcolor[rgb]{ 1,  1,  1}0.756 \\
    \bottomrule
    \end{tabular}%
  \label{tab:WithDNN}%
\end{table}%

\subsection{RQ2: Superiority} 
\textbf{Goal and Setup}. To verify the superiority of \texttt{RoMaC}, we construct an evaluation framework consisting of two dimensions:
1) the model’s adaptability to obfuscation-induced label noise; 2) the model’s robustness under varying ratios of code obfuscation.

% Table generated by Excel2LaTeX from sheet '总表新'
% Table generated by Excel2LaTeX from sheet '总表新'
% Table generated by Excel2LaTeX from sheet '总表新'
% Table generated by Excel2LaTeX from sheet '总表新'
\begin{table*}[htbp]
  \centering
            \tabcolsep = 0.10cm % 修改列间距
  \renewcommand{\arraystretch}{0.8} %行距
  \caption{Performance of RoMaC and other comparison methods.}
    \begin{tabular}{cccccccccccccccccccccc}
    \toprule
    \multicolumn{1}{c}{\multirow{3}[5]{*}{ObType}} & \multicolumn{1}{c}{\multirow{3}[5]{*}{R}} & \multicolumn{4}{c}{RoMaC}     & \multicolumn{4}{c}{DNN}       & \multicolumn{3}{c}{LDAM} &       & \multicolumn{4}{c}{MentorMix} & \multicolumn{4}{c}{MORSE} \\
\cmidrule{3-22}          &       & \multicolumn{2}{c}{Full classes} & \multicolumn{2}{c}{Tail classes} & \multicolumn{2}{c}{Full classes} & \multicolumn{2}{c}{Tail classes} & \multicolumn{2}{c}{Full classes} & \multicolumn{2}{c}{Tail classes} & \multicolumn{2}{c}{Full classes} & \multicolumn{2}{c}{Tail classes} & \multicolumn{2}{c}{Full classes} & \multicolumn{2}{c}{Tail classes} \\
\cmidrule{3-22}          &       & M-F1  & ACC   & M-F1  & ACC   & M-F1  & ACC   & M-F1  & ACC   & M-F1  & ACC   & M-F1  & ACC   & M-F1  & ACC   & M-F1  & ACC   & M-F1  & ACC   & M-F1  & ACC \\
    \midrule
    \multirow{5}[0]{*}{CSE} & 0.3   & \cellcolor[rgb]{ .929,  .945,  .976}\textbf{0.803} & \textbf{0.871} & \cellcolor[rgb]{ .929,  .945,  .976}\textbf{0.672} & \textbf{0.784} & \cellcolor[rgb]{ .929,  .945,  .976}0.645 & 0.772 & \cellcolor[rgb]{ .929,  .945,  .976}0.509 & 0.651 & \cellcolor[rgb]{ .929,  .945,  .976}0.641 & 0.763 & \cellcolor[rgb]{ .929,  .945,  .976}0.510 & 0.649 & \cellcolor[rgb]{ .929,  .945,  .976}0.619 & 0.752 & \cellcolor[rgb]{ .929,  .945,  .976}0.461 & 0.608 & \cellcolor[rgb]{ .929,  .945,  .976}0.724 & 0.809 & \cellcolor[rgb]{ .929,  .945,  .976}0.592 & 0.698 \\
          & 0.4   & \cellcolor[rgb]{ .929,  .945,  .976}\textbf{0.698} & \textbf{0.842} & \cellcolor[rgb]{ .929,  .945,  .976}\textbf{0.648} & \textbf{0.718} & \cellcolor[rgb]{ .929,  .945,  .976}0.580 & 0.793 & \cellcolor[rgb]{ .929,  .945,  .976}0.486 & 0.613 & \cellcolor[rgb]{ .929,  .945,  .976}0.517 & 0.737 & \cellcolor[rgb]{ .929,  .945,  .976}0.389 & 0.604 & \cellcolor[rgb]{ .929,  .945,  .976}0.527 & 0.715 & \cellcolor[rgb]{ .929,  .945,  .976}0.352 & 0.577 & \cellcolor[rgb]{ .929,  .945,  .976}0.646 & 0.771 & \cellcolor[rgb]{ .929,  .945,  .976}0.559 & 0.661 \\
          & 0.5   & \cellcolor[rgb]{ .929,  .945,  .976}\textbf{0.671} & \textbf{0.794} & \cellcolor[rgb]{ .929,  .945,  .976}\textbf{0.582} & \textbf{0.672} & \cellcolor[rgb]{ .929,  .945,  .976}0.503 & 0.735 & \cellcolor[rgb]{ .929,  .945,  .976}0.432 & 0.589 & \cellcolor[rgb]{ .929,  .945,  .976}0.484 & 0.694 & \cellcolor[rgb]{ .929,  .945,  .976}0.388 & 0.562 & \cellcolor[rgb]{ .929,  .945,  .976}0.516 & 0.683 & \cellcolor[rgb]{ .929,  .945,  .976}0.337 & 0.546 & \cellcolor[rgb]{ .929,  .945,  .976}0.607 & 0.723 & \cellcolor[rgb]{ .929,  .945,  .976}0.524 & 0.633 \\
          & 0.6   & \cellcolor[rgb]{ .929,  .945,  .976}\textbf{0.623} & \textbf{0.746} & \cellcolor[rgb]{ .929,  .945,  .976}\textbf{0.541} & \textbf{0.615} & \cellcolor[rgb]{ .929,  .945,  .976}0.458 & 0.662 & \cellcolor[rgb]{ .929,  .945,  .976}0.409 & 0.552 & \cellcolor[rgb]{ .929,  .945,  .976}0.423 & 0.650 & \cellcolor[rgb]{ .929,  .945,  .976}0.337 & 0.523 & \cellcolor[rgb]{ .929,  .945,  .976}0.420 & 0.648 & \cellcolor[rgb]{ .929,  .945,  .976}0.324 & 0.501 & \cellcolor[rgb]{ .929,  .945,  .976}0.563 & 0.692 & \cellcolor[rgb]{ .929,  .945,  .976}0.472 & 0.577 \\
          & 0.7   & \cellcolor[rgb]{ .929,  .945,  .976}\textbf{0.528} & \textbf{0.721} & \cellcolor[rgb]{ .929,  .945,  .976}\textbf{0.488} & \textbf{0.593} & \cellcolor[rgb]{ .929,  .945,  .976}0.442 & 0.615 & \cellcolor[rgb]{ .929,  .945,  .976}0.378 & 0.521 & \cellcolor[rgb]{ .929,  .945,  .976}0.397 & 0.609 & \cellcolor[rgb]{ .929,  .945,  .976}0.320 & 0.498 & \cellcolor[rgb]{ .929,  .945,  .976}0.318 & 0.611 & \cellcolor[rgb]{ .929,  .945,  .976}0.233 & 0.442 & \cellcolor[rgb]{ .929,  .945,  .976}0.462 & 0.659 & \cellcolor[rgb]{ .929,  .945,  .976}0.431 & 0.543 \\
              \midrule
    \multirow{5}[0]{*}{CRE} & 0.3   & \cellcolor[rgb]{ .929,  .945,  .976}\textbf{0.752} & \textbf{0.854} & \cellcolor[rgb]{ .929,  .945,  .976}\textbf{0.604} & \textbf{0.764} & \cellcolor[rgb]{ .929,  .945,  .976}0.677 & 0.783 & \cellcolor[rgb]{ .929,  .945,  .976}0.533 & 0.660 & \cellcolor[rgb]{ .929,  .945,  .976}0.678 & 0.771 & \cellcolor[rgb]{ .929,  .945,  .976}0.512 & 0.645 & \cellcolor[rgb]{ .929,  .945,  .976}0.619 & 0.758 & \cellcolor[rgb]{ .929,  .945,  .976}0.438 & 0.594 & \cellcolor[rgb]{ .929,  .945,  .976}0.706 & 0.785 & \cellcolor[rgb]{ .929,  .945,  .976}0.562 & 0.686 \\
          & 0.4   & \cellcolor[rgb]{ .929,  .945,  .976}\textbf{0.706} & \textbf{0.824} & \cellcolor[rgb]{ .929,  .945,  .976}\textbf{0.587} & \textbf{0.723} & \cellcolor[rgb]{ .929,  .945,  .976}0.610 & 0.736 & \cellcolor[rgb]{ .929,  .945,  .976}0.456 & 0.609 & \cellcolor[rgb]{ .929,  .945,  .976}0.632 & 0.741 & \cellcolor[rgb]{ .929,  .945,  .976}0.478 & 0.612 & \cellcolor[rgb]{ .929,  .945,  .976}0.587 & 0.715 & \cellcolor[rgb]{ .929,  .945,  .976}0.415 & 0.558 & \cellcolor[rgb]{ .929,  .945,  .976}0.672 & 0.742 & \cellcolor[rgb]{ .929,  .945,  .976}0.543 & 0.653 \\
          & 0.5   & \cellcolor[rgb]{ .929,  .945,  .976}\textbf{0.673} & \textbf{0.786} & \cellcolor[rgb]{ .929,  .945,  .976}\textbf{0.550} & \textbf{0.675} & \cellcolor[rgb]{ .929,  .945,  .976}0.597 & 0.712 & \cellcolor[rgb]{ .929,  .945,  .976}0.494 & 0.572 & \cellcolor[rgb]{ .929,  .945,  .976}0.611 & 0.693 & \cellcolor[rgb]{ .929,  .945,  .976}0.470 & 0.573 & \cellcolor[rgb]{ .929,  .945,  .976}0.555 & 0.686 & \cellcolor[rgb]{ .929,  .945,  .976}0.428 & 0.504 & \cellcolor[rgb]{ .929,  .945,  .976}0.641 & 0.710 & \cellcolor[rgb]{ .929,  .945,  .976}0.521 & 0.631 \\
          & 0.6   & \cellcolor[rgb]{ .929,  .945,  .976}\textbf{0.653} & \textbf{0.762} & \cellcolor[rgb]{ .929,  .945,  .976}\textbf{0.550} & \textbf{0.648} & \cellcolor[rgb]{ .929,  .945,  .976}0.559 & 0.668 & \cellcolor[rgb]{ .929,  .945,  .976}0.432 & 0.561 & \cellcolor[rgb]{ .929,  .945,  .976}0.558 & 0.649 & \cellcolor[rgb]{ .929,  .945,  .976}0.405 & 0.558 & \cellcolor[rgb]{ .929,  .945,  .976}0.495 & 0.628 & \cellcolor[rgb]{ .929,  .945,  .976}0.337 & 0.460 & \cellcolor[rgb]{ .929,  .945,  .976}0.614 & 0.689 & \cellcolor[rgb]{ .929,  .945,  .976}0.514 & 0.595 \\
          & 0.7   & \cellcolor[rgb]{ .929,  .945,  .976}\textbf{0.643} & \textbf{0.745} & \cellcolor[rgb]{ .929,  .945,  .976}\textbf{0.522} & \textbf{0.605} & \cellcolor[rgb]{ .929,  .945,  .976}0.520 & 0.612 & \cellcolor[rgb]{ .929,  .945,  .976}0.410 & 0.546 & \cellcolor[rgb]{ .929,  .945,  .976}0.539 & 0.608 & \cellcolor[rgb]{ .929,  .945,  .976}0.399 & 0.530 & \cellcolor[rgb]{ .929,  .945,  .976}0.503 & 0.605 & \cellcolor[rgb]{ .929,  .945,  .976}0.351 & 0.451 & \cellcolor[rgb]{ .929,  .945,  .976}0.609 & 0.672 & \cellcolor[rgb]{ .929,  .945,  .976}0.490 & 0.560 \\
              \midrule
    \multirow{5}[0]{*}{CID} & 0.3   & \cellcolor[rgb]{ .929,  .945,  .976}\textbf{0.769} & \textbf{0.856} & \cellcolor[rgb]{ .929,  .945,  .976}\textbf{0.624} & \textbf{0.775} & \cellcolor[rgb]{ .929,  .945,  .976}0.649 & 0.758 & \cellcolor[rgb]{ .929,  .945,  .976}0.522 & 0.649 & \cellcolor[rgb]{ .929,  .945,  .976}0.663 & 0.754 & \cellcolor[rgb]{ .929,  .945,  .976}0.514 & 0.651 & \cellcolor[rgb]{ .929,  .945,  .976}0.627 & 0.762 & \cellcolor[rgb]{ .929,  .945,  .976}0.443 & 0.589 & \cellcolor[rgb]{ .929,  .945,  .976}0.718 & 0.783 & \cellcolor[rgb]{ .929,  .945,  .976}0.599 & 0.693 \\
          & 0.4   & \cellcolor[rgb]{ .929,  .945,  .976}\textbf{0.728} & \textbf{0.813} & \cellcolor[rgb]{ .929,  .945,  .976}\textbf{0.587} & \textbf{0.744} & \cellcolor[rgb]{ .929,  .945,  .976}0.604 & 0.716 & \cellcolor[rgb]{ .929,  .945,  .976}0.473 & 0.612 & \cellcolor[rgb]{ .929,  .945,  .976}0.625 & 0.696 & \cellcolor[rgb]{ .929,  .945,  .976}0.470 & 0.619 & \cellcolor[rgb]{ .929,  .945,  .976}0.615 & 0.740 & \cellcolor[rgb]{ .929,  .945,  .976}0.459 & 0.572 & \cellcolor[rgb]{ .929,  .945,  .976}0.691 & 0.754 & \cellcolor[rgb]{ .929,  .945,  .976}0.568 & 0.663 \\
          & 0.5   & \cellcolor[rgb]{ .929,  .945,  .976}\textbf{0.654} & \textbf{0.794} & \cellcolor[rgb]{ .929,  .945,  .976}\textbf{0.527} & \textbf{0.699} & \cellcolor[rgb]{ .929,  .945,  .976}0.588 & 0.685 & \cellcolor[rgb]{ .929,  .945,  .976}0.472 & 0.586 & \cellcolor[rgb]{ .929,  .945,  .976}0.558 & 0.626 & \cellcolor[rgb]{ .929,  .945,  .976}0.414 & 0.581 & \cellcolor[rgb]{ .929,  .945,  .976}0.540 & 0.705 & \cellcolor[rgb]{ .929,  .945,  .976}0.392 & 0.532 & \cellcolor[rgb]{ .929,  .945,  .976}0.626 & 0.731 & \cellcolor[rgb]{ .929,  .945,  .976}0.517 & 0.641 \\
          & 0.6   & \cellcolor[rgb]{ .929,  .945,  .976}\textbf{0.665} & \textbf{0.759} & \cellcolor[rgb]{ .929,  .945,  .976}\textbf{0.541} & \textbf{0.652} & \cellcolor[rgb]{ .929,  .945,  .976}0.552 & 0.635 & \cellcolor[rgb]{ .929,  .945,  .976}0.482 & 0.568 & \cellcolor[rgb]{ .929,  .945,  .976}0.546 & 0.593 & \cellcolor[rgb]{ .929,  .945,  .976}0.428 & 0.546 & \cellcolor[rgb]{ .929,  .945,  .976}0.505 & 0.659 & \cellcolor[rgb]{ .929,  .945,  .976}0.379 & 0.499 & \cellcolor[rgb]{ .929,  .945,  .976}0.622 & 0.712 & \cellcolor[rgb]{ .929,  .945,  .976}0.526 & 0.598 \\
          & 0.7   & \cellcolor[rgb]{ .929,  .945,  .976}\textbf{0.640} & \textbf{0.736} & \cellcolor[rgb]{ .929,  .945,  .976}\textbf{0.526} & \textbf{0.613} & \cellcolor[rgb]{ .929,  .945,  .976}0.513 & 0.607 & \cellcolor[rgb]{ .929,  .945,  .976}0.407 & 0.550 & \cellcolor[rgb]{ .929,  .945,  .976}0.516 & 0.571 & \cellcolor[rgb]{ .929,  .945,  .976}0.406 & 0.504 & \cellcolor[rgb]{ .929,  .945,  .976}0.510 & 0.613 & \cellcolor[rgb]{ .929,  .945,  .976}0.373 & 0.473 & \cellcolor[rgb]{ .929,  .945,  .976}0.608 & 0.680 & \cellcolor[rgb]{ .929,  .945,  .976}0.475 & 0.556 \\
              \midrule
    \multirow{5}[1]{*}{MIX} & 0.3   & \cellcolor[rgb]{ .929,  .945,  .976}\textbf{0.755} & \textbf{0.835} & \cellcolor[rgb]{ .929,  .945,  .976}\textbf{0.639} & \textbf{0.759} & \cellcolor[rgb]{ .929,  .945,  .976}0.649 & 0.754 & \cellcolor[rgb]{ .929,  .945,  .976}0.550 & 0.654 & \cellcolor[rgb]{ .929,  .945,  .976}0.696 & 0.758 & \cellcolor[rgb]{ .929,  .945,  .976}0.554 & 0.655 & \cellcolor[rgb]{ .929,  .945,  .976}0.680 & 0.771 & \cellcolor[rgb]{ .929,  .945,  .976}0.529 & 0.618 & \cellcolor[rgb]{ .929,  .945,  .976}0.713 & 0.765 & \cellcolor[rgb]{ .929,  .945,  .976}0.622 & 0.691 \\
          & 0.4   & \cellcolor[rgb]{ .929,  .945,  .976}\textbf{0.727} & \textbf{0.813} & \cellcolor[rgb]{ .929,  .945,  .976}\textbf{0.625} & \textbf{0.721} & \cellcolor[rgb]{ .929,  .945,  .976}0.609 & 0.722 & \cellcolor[rgb]{ .929,  .945,  .976}0.504 & 0.628 & \cellcolor[rgb]{ .929,  .945,  .976}0.648 & 0.719 & \cellcolor[rgb]{ .929,  .945,  .976}0.510 & 0.618 & \cellcolor[rgb]{ .929,  .945,  .976}0.651 & 0.729 & \cellcolor[rgb]{ .929,  .945,  .976}0.550 & 0.583 & \cellcolor[rgb]{ .929,  .945,  .976}0.671 & 0.750 & \cellcolor[rgb]{ .929,  .945,  .976}0.607 & 0.670 \\
          & 0.5   & \cellcolor[rgb]{ .929,  .945,  .976}\textbf{0.705} & \textbf{0.791} & \cellcolor[rgb]{ .929,  .945,  .976}\textbf{0.617} & \textbf{0.696} & \cellcolor[rgb]{ .929,  .945,  .976}0.592 & 0.695 & \cellcolor[rgb]{ .929,  .945,  .976}0.515 & 0.605 & \cellcolor[rgb]{ .929,  .945,  .976}0.654 & 0.673 & \cellcolor[rgb]{ .929,  .945,  .976}0.540 & 0.590 & \cellcolor[rgb]{ .929,  .945,  .976}0.620 & 0.695 & \cellcolor[rgb]{ .929,  .945,  .976}0.443 & 0.558 & \cellcolor[rgb]{ .929,  .945,  .976}0.664 & 0.737 & \cellcolor[rgb]{ .929,  .945,  .976}0.592 & 0.623 \\
          & 0.6   & \cellcolor[rgb]{ .929,  .945,  .976}\textbf{0.700} & \textbf{0.776} & \cellcolor[rgb]{ .929,  .945,  .976}\textbf{0.604} & \textbf{0.657} & \cellcolor[rgb]{ .929,  .945,  .976}0.575 & 0.643 & \cellcolor[rgb]{ .929,  .945,  .976}0.509 & 0.588 & \cellcolor[rgb]{ .929,  .945,  .976}0.597 & 0.633 & \cellcolor[rgb]{ .929,  .945,  .976}0.522 & 0.563 & \cellcolor[rgb]{ .929,  .945,  .976}0.575 & 0.654 & \cellcolor[rgb]{ .929,  .945,  .976}0.453 & 0.544 & \cellcolor[rgb]{ .929,  .945,  .976}0.641 & 0.723 & \cellcolor[rgb]{ .929,  .945,  .976}0.562 & 0.594 \\
          & 0.7   & \cellcolor[rgb]{ .929,  .945,  .976}\textbf{0.685} & \textbf{0.752} & \cellcolor[rgb]{ .929,  .945,  .976}\textbf{0.593} & \textbf{0.631} & \cellcolor[rgb]{ .929,  .945,  .976}0.558 & 0.618 & \cellcolor[rgb]{ .929,  .945,  .976}0.497 & 0.573 & \cellcolor[rgb]{ .929,  .945,  .976}0.603 & 0.581 & \cellcolor[rgb]{ .929,  .945,  .976}0.510 & 0.558 & \cellcolor[rgb]{ .929,  .945,  .976}0.545 & 0.628 & \cellcolor[rgb]{ .929,  .945,  .976}0.447 & 0.532 & \cellcolor[rgb]{ .929,  .945,  .976}0.632 & 0.691 & \cellcolor[rgb]{ .929,  .945,  .976}0.553 & 0.563 \\
    \bottomrule
    \end{tabular}%
  \label{tab:Sumtable}%
\end{table*}%

\textbf{Analysis and Result}. Table \ref{tab:Sumtable} compares \texttt{RoMaC} with various methods under different conditions. The left columns indicate the obfuscation type of the synthetic dataset, with "R" representing the specific obfuscation noise ratio within the dataset, which means the proportion of obfuscated samples. The table presents results for \texttt{RoMaC}, DNN, LDAM (a method for class imbalance), MentorMix (a method for label noise), and the state-of-the-art method MORSE in this domain. 

Overall, across four different obfuscation types, \texttt{RoMaC} achieves the best performance, demonstrating its superiority in malware family classification under code obfuscation. At the obfuscation-induced noise ratio of 0.3, \texttt{RoMaC} achieves optimal performance on CSE-obfuscated data, with the four metrics reaching 0.803, 0.871, 0.672, and 0.784. A gradual increase in the obfuscation-induced noise ratio results in a significant decline in performance, particularly on CSE-obfuscated data. Similar trends are observed in other comparison methods, which may be attributed to the model's overfitting to label noise. At a higher obfuscation noise ratio, the model under MIX obfuscation performs best, possibly because singular obfuscation-induced noise helps the model capture systematic error patterns (e.g., unique label alterations of specific features caused by CSE). In contrast, mixed obfuscation introduces diverse label noise, forcing the model to learn more robust features. This diversity serves as implicit regularization, mitigating model overfitting under high-noise conditions and improving generalization capabilities.

Three key phenomena can be observed when comparing with other methods.

1) In certain scenarios, existing class imbalance learning methods perform worse than standard DNN. This is because these methods fail to distinguish noisy samples, and their erroneous focus on these samples further undermines the model's generalizability.

2) Compared with existing methods, \texttt{RoMaC} demonstrates significantly better performance in addressing class imbalance. For instance, \texttt{RoMaC} shows significant improvements in tail-class performance metrics. Specifically, under the CSE obfuscation setting with an obfuscation-induced noise ratio of 0.3, \texttt{RoMaC} outperforms the SOTA method MORSE by 8.0\% and 8.6\%, and the standard DNN baseline by 16.3\% and 13.3\% in terms of the performance in tail classes. This further underscores its enhanced capability in addressing class imbalance. Notably, \texttt{RoMaC} maintains strong performance even as obfuscation noise increases. Under CRE obfuscation, when the noise ratio ranges from 0.3 to 0.7, \texttt{RoMaC} exhibits only a 10.9\% drop in overall M-F1 (from 0.752 to 0.643), whereas LAMD drops by 13.9\%. This demonstrates the superior robustness of \texttt{RoMaC} under challenging, noise-prone conditions.

3) Compared to methods that focus exclusively on either class imbalance or label noise, the advantages of existing methods targeting both class imbalance and label noise are not pronounced. This limitation stems from their failure to consider the interplay between class imbalance and label noise.

\begin{figure}[htbp]
  \centering
  \includegraphics[width=0.45\textwidth]{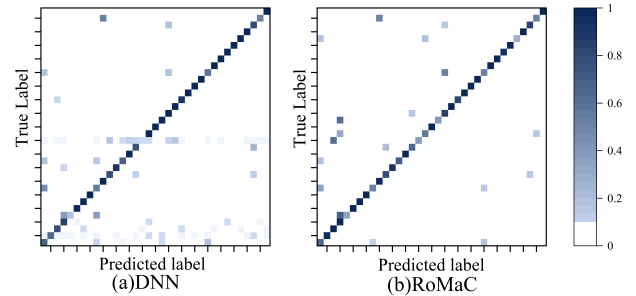} % 调整宽度并指定文件名
  \caption{Confusion matrix for DNN and \texttt{RoMaC}.}
  \label{fig:CM}
\end{figure}

We further demonstrate the superior performance of \texttt{RoMaC} by visually presenting the results through a confusion matrix, as shown in Fig. \ref{fig:CM}, which compares the performance of DNN and \texttt{RoMaC}. In the confusion matrix, the value located at the $i$-th row and $j$-th column represents the proportion of samples from the $i$-th class predicted as the $j$-th class. Cells with higher proportions are assigned darker blue shades. We observe that the DNN frequently misclassifies a significant portion of tail-class samples. For example, label-16 samples create a "light-colored cell line" in the DNN confusion matrix, indicating severe misclassification and poor recognition capability for this class. 

\begin{figure}[htbp]
  \centering
  \includegraphics[width=0.45\textwidth]{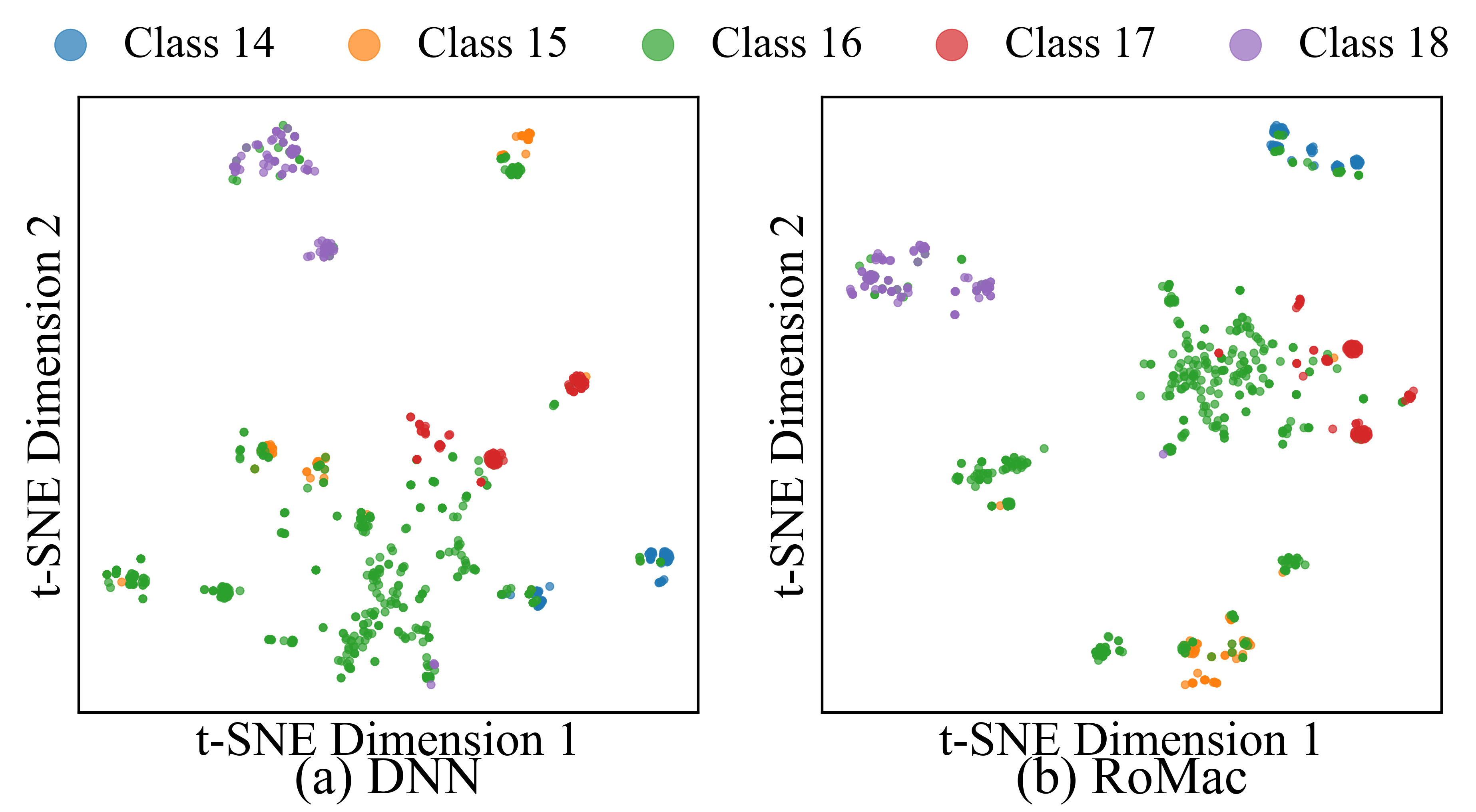} % 调整宽度并指定文件名
  \caption{t-SNE results for DNN and \texttt{RoMaC}.}
  \label{fig:TSNE}
\end{figure}

To further analyze the feature representations, we utilize t-SNE (t-distributed stochastic neighbor embedding) for dimensionality reduction, visualizing the features extracted from the penultimate layer of both the DNN and \texttt{RoMaC}'s network. The results are presented in Figure \ref{fig:TSNE}. After applying \texttt{RoMaC}, the label-16 samples demonstrate greater similarity and compactness in the feature space, thus validating the DNN's difficulties with tail data and highlighting the effectiveness of \texttt{RoMaC} in addressing the class imbalance.

% In addition to label noise caused by code obfuscation, other sources such as random errors introduced by automated labeling platforms also contribute to label noise. To model these scenarios, we incorporate random noise alongside obfuscation-induced noise (see Appendix B for details).

\subsection{RQ3: Hyperparameter Analysis }\label{RQ3}
\textbf{Goal and Setup}. To achieve the best performance, we conduct experiments to analyze the following hyperparameters: 1) the weak regularization weight $\lambda_{weak}$; 2) the number of ensemble models $n$ and the reweighting coefficient $\mu$; 3) the proportion of unlabeled data $d$. We conduct our experiments under the CSE obfuscation.

\textbf{Analysis and Result}. 
Our experimental results and analysis are given below. 

\begin{figure}[htbp]
  \centering
  \includegraphics[width=0.45\textwidth]{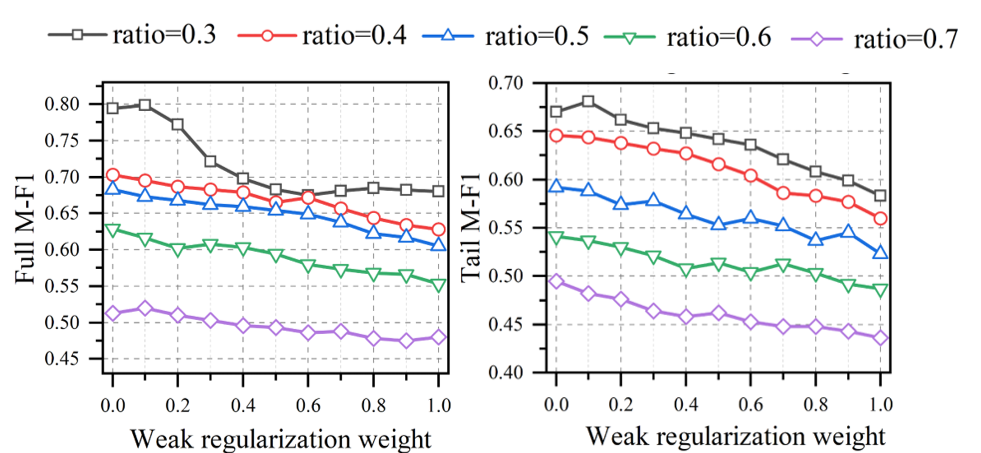} % 调整宽度并指定文件名
  \caption{Effect of weak regularization weight  $\lambda_{weak}$.}
  \label{fig:DC}
\end{figure}

\textit{Weak regularization weight $\lambda_{weak}$}: Fig. \ref{fig:DC} illustrates the impact of $\lambda_{weak}$ on model performance. The figure indicates a downward trend in overall model performance as $\lambda_{weak}$ increases from 0 to 1. 
% The model exhibits strong classification effectiveness when $\lambda_{weak}$ is between 0 and 0.2. Most configurations achieve peak performance at $\lambda_{weak} = 0$, with some optimal results occurring at $\lambda_{weak}$ = 0.1. 
This observation supports our hypothesis that tail-class samples, whose pseudo-labels are unreliable due to class imbalance, adversely affect model performance when included in training. Lower values of $\lambda_{weak}$ significantly mitigate the impact of misclassification on tail data while preserving sample diversity, thus enhancing overall performance. Therefore, we conservatively set $\lambda_{weak}$ to 0.1.

\textit{The number of ensemble models $n$ and reweighting coefficient $\mu$}. To optimize ensemble reweighting, we establish two objectives: 1) Determine the maximum $\mu$ value - Excessive $\mu$ values can lead to model bias towards tail data, rendering such models ineffective for the ensemble, ultimately degrading ensemble performance and increasing computational costs. 2) Find the appropriate number of models in the pool - After establishing the maximum $\mu$ value, we aim to identify the optimal number of models in the pool to ensure that $\mu$ values span from "complete head bias" to "balanced". Indiscriminate increases in the number of models do not enhance ensemble performance and instead incur unnecessary computational overhead. 

In model ensembling, diversity is a critical factor. The performance improvements of ensemble models primarily stem from the predictive divergence among base learners, rather than solely relying on the individual accuracy of each model \cite{ortega2022diversity}. When high-performing base learners exhibit low diversity, their predictions tend to be overly similar, thereby constraining the ensemble's potential gains \cite{shen2022divbo}. Therefore, we advocate evaluating ensemble effectiveness from the perspective of model diversity. If incorporating additional models fails to increase the diversity, their inclusion is unlikely to yield substantial performance benefits. We choose Kohavi-Wolpert Variance ($KW$) to assess ensemble model diversity \cite{KW} by quantifying the prediction disagreement among base learners for each sample.
% $$
% \mathrm{KW}=\frac{1}{N \cdot M^2} \sum_{i=1}^N k_i\left(M-k_i\right)
% \label{eqa:KW}
% $$
\begin{equation}
K W=\frac{1}{N M^2} \sum_{i=1}^N k_i\left(M-k_i\right). \quad
\label{eqa:KW}
\end{equation}
\begin{equation}
k_i=\sum_{m=1}^M \mathbb{I}\left(f_m\left(x_i\right)=y_i\right).
\label{eqa:KW1}
\end{equation}

In Equation \ref{eqa:KW} and \ref{eqa:KW1}, $N$ represents the total number of samples in the test set, $M$ denotes the number of base models in ensemble learning, and $k_i$ indicates the number of times the $i$-th sample is correctly predicted by the base model ($0 \le k_i \le M$). When $0.1 \le KW \le 0.15$, there is some diversity among the models. The ensemble enhancement becomes pronounced.

\begin{figure}[htbp]
  \centering
  \includegraphics{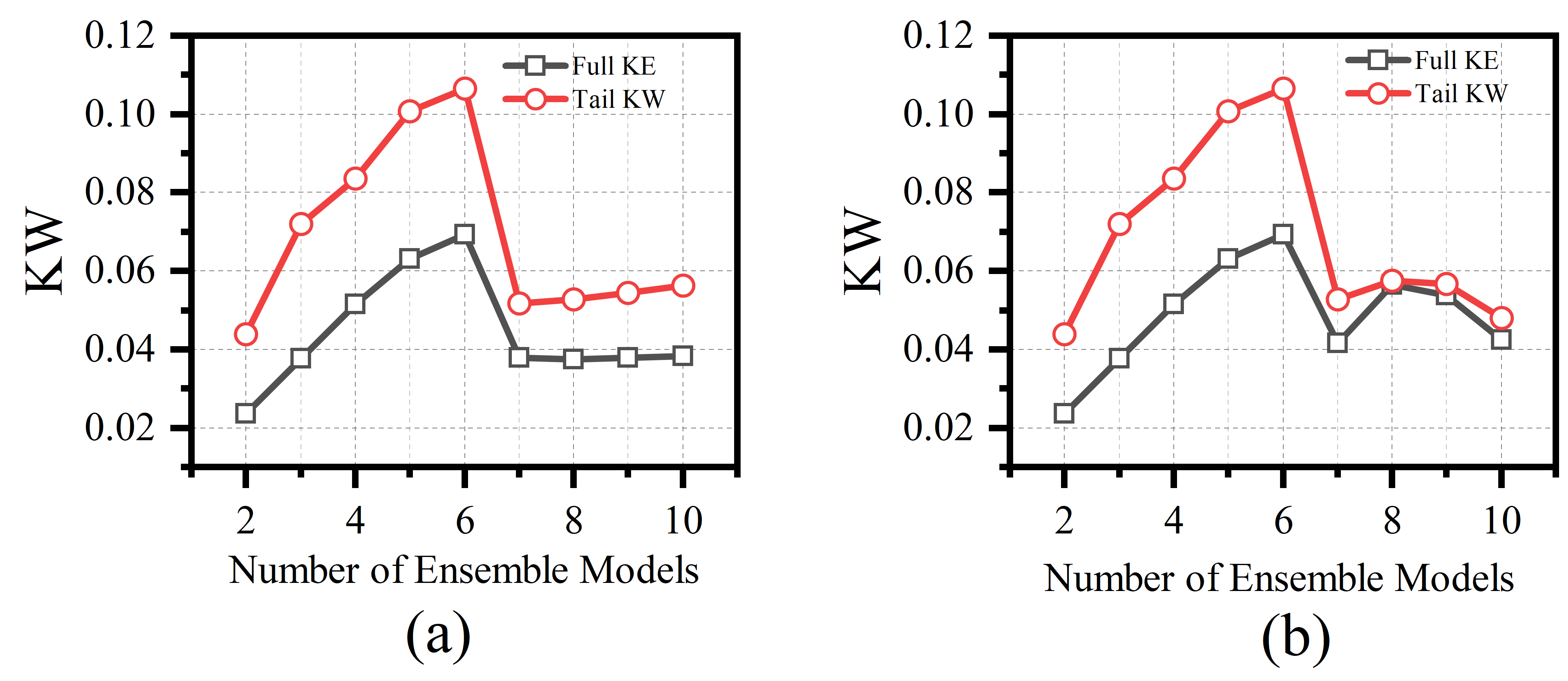} % 调整宽度并指定文件名
  \caption{The $KW$ of ensemble models: (a) Determine the maximum $\mu$ value, (b) Find the appropriate number of models in the pool.  }
  \label{fig:ensemble}
\end{figure}

Fig. \ref{fig:ensemble} illustrates how the number of models influences model diversity. The left subplot explores the effect of gradually increasing the maximum reweighting coefficient $\mu$, starting from a base ensemble of two models (a non-reweighted model and a model with $\mu=0.1$), and incrementing $\mu$ by 0.2 for each additional model. The right subplot analyzes how diversity changes with varying model counts, keeping the maximum $\mu$ fixed. From the figure, we observe that model diversity peaks when the number of models is 6. Increasing or decreasing the number of models from this point leads to a noticeable decline in diversity. This diversity contributes to better ensemble performance by reducing overfitting and enhancing generalization. Importantly, the higher KW of tail classes indicates strong complementarity for minority classes, enabling the ensemble to better capture tail patterns and reduce class imbalance. 
Consequently, we select the model pool $[non-reweighted, \mu=0.2, 0.4, 0.6, 0.8, 1]$ for the ensemble, striking an optimal balance between diversity and computational efficiency. This result theoretically supports the observed performance gains of \texttt{RoMaC}, confirming the effectiveness of the ensemble reweighting in handling class imbalance and label noise.

\textit{The unlabeled data proportion $d$}.  We investigate the impact of unlabeled sample proportions on model performance. Considering that our dataset's maximum obfuscation-induced label noise ratio is 70\%, we set the range of $d$ from 0 to 0.5.

% Table generated by Excel2LaTeX from sheet '超参数d	d							'
% Table generated by Excel2LaTeX from sheet '超参数d	d							'
\begin{table}[htbp]\footnotesize
  \centering
          \tabcolsep = 0.05cm % 修改列间距
\renewcommand{\arraystretch}{0.7} %行距
  \caption{The performance of \texttt{RoMaC} with different $d$.}
    \begin{tabular}{c|c|ccccccc}
    \toprule
    \multicolumn{1}{l|}{Parameter} & \multicolumn{1}{p{7.69em}|}{\cellcolor[rgb]{ .949,  .949,  .949}\diagbox[width=7em]{R}{$d$}} & \cellcolor[rgb]{ .949,  .949,  .949}0.05 & \cellcolor[rgb]{ .949,  .949,  .949}0.1 & \cellcolor[rgb]{ .949,  .949,  .949}0.15 & \cellcolor[rgb]{ .949,  .949,  .949}0.2 & \cellcolor[rgb]{ .949,  .949,  .949}0.3 & \cellcolor[rgb]{ .949,  .949,  .949}0.4 & \cellcolor[rgb]{ .949,  .949,  .949}0.5 \\
    \midrule
    \multirow{5}[2]{*}{Full M-F1} & 0.3   & \cellcolor[rgb]{ .522,  .663,  .839}0.774 & \cellcolor[rgb]{ .671,  .765,  .89}0.803 & \cellcolor[rgb]{ .596,  .714,  .863}0.742 & \cellcolor[rgb]{ .663,  .761,  .886}0.712 & \cellcolor[rgb]{ .761,  .831,  .922}0.669 & \cellcolor[rgb]{ .91,  .933,  .973}0.605 & \cellcolor[rgb]{ .98,  .784,  .792}0.458 \\
          & 0.4   & \cellcolor[rgb]{ .706,  .792,  .902}0.693 & \cellcolor[rgb]{ .984,  .886,  .898}0.698 & \cellcolor[rgb]{ .659,  .757,  .886}0.715 & \cellcolor[rgb]{ .71,  .792,  .902}0.692 & \cellcolor[rgb]{ .82,  .871,  .941}0.644 & \cellcolor[rgb]{ .988,  .988,  1}0.569 & \cellcolor[rgb]{ .98,  .737,  .749}0.433 \\
          & 0.5   & \cellcolor[rgb]{ .804,  .859,  .937}0.651 & \cellcolor[rgb]{ .98,  .812,  .82}0.671 & \cellcolor[rgb]{ .816,  .867,  .941}0.646 & \cellcolor[rgb]{ .867,  .902,  .957}0.624 & \cellcolor[rgb]{ .925,  .945,  .98}0.597 & \cellcolor[rgb]{ .984,  .957,  .969}0.552 & \cellcolor[rgb]{ .976,  .663,  .671}0.392 \\
          & 0.6   & \cellcolor[rgb]{ .871,  .906,  .961}0.622 & \cellcolor[rgb]{ .976,  .675,  .686}0.623 & \cellcolor[rgb]{ .867,  .902,  .957}0.624 & \cellcolor[rgb]{ .898,  .925,  .969}0.610 & \cellcolor[rgb]{ .984,  .914,  .925}0.530 & \cellcolor[rgb]{ .98,  .82,  .831}0.478 & \cellcolor[rgb]{ .976,  .588,  .596}0.351 \\
          & 0.7   & \cellcolor[rgb]{ .984,  .922,  .933}0.534 & \cellcolor[rgb]{ .973,  .412,  .42}0.528 & \cellcolor[rgb]{ .984,  .894,  .906}0.518 & \cellcolor[rgb]{ .984,  .867,  .878}0.504 & \cellcolor[rgb]{ .98,  .827,  .839}0.482 & \cellcolor[rgb]{ .98,  .757,  .765}0.443 & \cellcolor[rgb]{ .973,  .529,  .541}0.320 \\
    \midrule
    \multirow{5}[2]{*}{Full ACC} & 0.3   & \cellcolor[rgb]{ .353,  .541,  .776}0.848 & \cellcolor[rgb]{ .353,  .541,  .776}0.871 & \cellcolor[rgb]{ .38,  .561,  .788}0.837 & \cellcolor[rgb]{ .439,  .604,  .808}0.811 & \cellcolor[rgb]{ .643,  .745,  .878}0.721 & \cellcolor[rgb]{ .894,  .922,  .969}0.612 & \cellcolor[rgb]{ .984,  .941,  .953}0.544 \\
          & 0.4   & \cellcolor[rgb]{ .408,  .58,  .796}0.824 & \cellcolor[rgb]{ .49,  .639,  .827}0.842 & \cellcolor[rgb]{ .42,  .588,  .8}0.819 & \cellcolor[rgb]{ .498,  .643,  .827}0.786 & \cellcolor[rgb]{ .753,  .824,  .918}0.674 & \cellcolor[rgb]{ .984,  .965,  .976}0.557 & \cellcolor[rgb]{ .98,  .827,  .839}0.483 \\
          & 0.5   & \cellcolor[rgb]{ .49,  .639,  .827}0.789 & \cellcolor[rgb]{ .71,  .792,  .902}0.794 & \cellcolor[rgb]{ .486,  .635,  .824}0.791 & \cellcolor[rgb]{ .569,  .694,  .855}0.754 & \cellcolor[rgb]{ .882,  .914,  .965}0.616 & \cellcolor[rgb]{ .984,  .843,  .855}0.491 & \cellcolor[rgb]{ .976,  .659,  .671}0.391 \\
          & 0.6   & \cellcolor[rgb]{ .576,  .698,  .855}0.751 & \cellcolor[rgb]{ .933,  .949,  .98}0.746 & \cellcolor[rgb]{ .604,  .718,  .867}0.739 & \cellcolor[rgb]{ .663,  .761,  .886}0.713 & \cellcolor[rgb]{ .914,  .937,  .976}0.602 & \cellcolor[rgb]{ .98,  .796,  .804}0.464 & \cellcolor[rgb]{ .976,  .62,  .627}0.368 \\
          & 0.7   & \cellcolor[rgb]{ .682,  .773,  .894}0.704 & \cellcolor[rgb]{ .984,  .953,  .961}0.721 & \cellcolor[rgb]{ .702,  .788,  .902}0.695 & \cellcolor[rgb]{ .741,  .816,  .914}0.678 & \cellcolor[rgb]{ .988,  .988,  1}0.569 & \cellcolor[rgb]{ .976,  .671,  .678}0.396 & \cellcolor[rgb]{ .976,  .573,  .58}0.343 \\
    \midrule
    \multirow{5}[2]{*}{Tail M-F1} & 0.3   & \cellcolor[rgb]{ .827,  .875,  .945}0.640 & \cellcolor[rgb]{ .757,  .824,  .918}0.672 & \cellcolor[rgb]{ .773,  .839,  .925}0.664 & \cellcolor[rgb]{ .824,  .871,  .941}0.643 & \cellcolor[rgb]{ .914,  .937,  .976}0.602 & \cellcolor[rgb]{ .984,  .976,  .988}0.564 & \cellcolor[rgb]{ .984,  .855,  .867}0.497 \\
          & 0.4   & \cellcolor[rgb]{ .961,  .969,  .992}0.582 & \cellcolor[rgb]{ .812,  .863,  .937}0.648 & \cellcolor[rgb]{ .851,  .894,  .953}0.630 & \cellcolor[rgb]{ .898,  .925,  .969}0.610 & \cellcolor[rgb]{ .98,  .98,  .996}0.574 & \cellcolor[rgb]{ .984,  .945,  .957}0.546 & \cellcolor[rgb]{ .98,  .773,  .784}0.453 \\
          & 0.5   & \cellcolor[rgb]{ .984,  .965,  .976}0.558 & \cellcolor[rgb]{ .961,  .969,  .992}0.582 & \cellcolor[rgb]{ .984,  .98,  .992}0.566 & \cellcolor[rgb]{ .984,  .941,  .953}0.545 & \cellcolor[rgb]{ .984,  .937,  .949}0.543 & \cellcolor[rgb]{ .984,  .882,  .894}0.512 & \cellcolor[rgb]{ .98,  .714,  .722}0.419 \\
          & 0.6   & \cellcolor[rgb]{ .984,  .937,  .949}0.542 & \cellcolor[rgb]{ .984,  .933,  .945}0.541 & \cellcolor[rgb]{ .984,  .941,  .953}0.545 & \cellcolor[rgb]{ .984,  .882,  .894}0.512 & \cellcolor[rgb]{ .984,  .875,  .882}0.507 & \cellcolor[rgb]{ .976,  .698,  .706}0.411 & \cellcolor[rgb]{ .973,  .522,  .529}0.315 \\
          & 0.7   & \cellcolor[rgb]{ .984,  .851,  .859}0.494 & \cellcolor[rgb]{ .98,  .839,  .847}0.488 & \cellcolor[rgb]{ .984,  .918,  .929}0.532 & \cellcolor[rgb]{ .98,  .835,  .847}0.486 & \cellcolor[rgb]{ .98,  .765,  .776}0.448 & \cellcolor[rgb]{ .976,  .608,  .616}0.362 & \cellcolor[rgb]{ .973,  .435,  .443}0.268 \\
    \midrule
    \multirow{5}[2]{*}{Tail ACC} & 0.3   & \cellcolor[rgb]{ .596,  .714,  .863}0.742 & \cellcolor[rgb]{ .502,  .647,  .831}0.784 & \cellcolor[rgb]{ .557,  .686,  .851}0.759 & \cellcolor[rgb]{ .702,  .788,  .902}0.695 & \cellcolor[rgb]{ .839,  .886,  .949}0.635 & \cellcolor[rgb]{ .984,  .961,  .973}0.556 & \cellcolor[rgb]{ .976,  .663,  .675}0.393 \\
          & 0.4   & \cellcolor[rgb]{ .733,  .812,  .914}0.681 & \cellcolor[rgb]{ .651,  .753,  .882}0.718 & \cellcolor[rgb]{ .706,  .788,  .902}0.694 & \cellcolor[rgb]{ .89,  .922,  .969}0.613 & \cellcolor[rgb]{ .984,  .973,  .984}0.561 & \cellcolor[rgb]{ .984,  .906,  .914}0.524 & \cellcolor[rgb]{ .976,  .6,  .608}0.358 \\
          & 0.5   & \cellcolor[rgb]{ .824,  .875,  .945}0.642 & \cellcolor[rgb]{ .757,  .824,  .918}0.672 & \cellcolor[rgb]{ .733,  .808,  .91}0.682 & \cellcolor[rgb]{ .988,  .988,  1}0.569 & \cellcolor[rgb]{ .984,  .855,  .867}0.497 & \cellcolor[rgb]{ .98,  .78,  .792}0.457 & \cellcolor[rgb]{ .976,  .561,  .569}0.336 \\
          & 0.6   & \cellcolor[rgb]{ .918,  .937,  .976}0.601 & \cellcolor[rgb]{ .886,  .918,  .965}0.615 & \cellcolor[rgb]{ .937,  .953,  .984}0.593 & \cellcolor[rgb]{ .984,  .894,  .906}0.518 & \cellcolor[rgb]{ .98,  .776,  .784}0.454 & \cellcolor[rgb]{ .976,  .639,  .651}0.380 & \cellcolor[rgb]{ .973,  .459,  .467}0.281 \\
          & 0.7   & \cellcolor[rgb]{ .984,  .953,  .965}0.550 & \cellcolor[rgb]{ .937,  .953,  .984}0.593 & \cellcolor[rgb]{ .945,  .961,  .988}0.588 & \cellcolor[rgb]{ .984,  .851,  .863}0.496 & \cellcolor[rgb]{ .976,  .655,  .663}0.388 & \cellcolor[rgb]{ .976,  .616,  .627}0.367 & \cellcolor[rgb]{ .973,  .412,  .42}0.254 \\
    \bottomrule

    \end{tabular}%
  \label{tab:unlabel d}%
\end{table}%

Table \ref{tab:unlabel d} demonstrates the impact of the unlabeled data proportion $d$ on \texttt{RoMaC}. R indicates the obfuscation-induced noise ratio. The results reveal that model performance peaks when $d$ ranges between 0.05 and 0.15 under all conditions. As $d$ increases, model performance fluctuates within an acceptable range until it reaches a threshold. When  $d$  exceeds this threshold, model performance deteriorates significantly. For instance, with an obfuscation-induced noise ratio of 0.3, the threshold for $d$ is approximately 0.2. Beyond this threshold, an increasing amount of clean data is misclassified as noise, alongside actual noisy data. This misclassification-induced reduction in clean data undermines the effectiveness of the reweighting component. Considering both the need for performance optimization and the practical necessity to prevent the misclassification of clean data, we conservatively set $d$ = 0.1 for all experiments.

\subsection{Ablation Study}
\textbf{Goal and Setup}. To understand the role of each component in \texttt{RoMaC}, we conduct ablation study through starting with a standard DNN and incrementally adding each component to evaluate its impact on performance under a CSE obfuscation-induced noise ratio of 0.3. 

% Table generated by Excel2LaTeX from sheet '消融实验'
\begin{table}[htbp]
  \centering
  \tabcolsep = 0.08cm % 修改列间距
  \caption{Ablation Study.}
    \begin{tabular}{l|cc|cc}
    \toprule
    \multicolumn{1}{c|}{\multirow{2}[4]{*}{Experiment Setting}} & \multicolumn{2}{c|}{Full classes} & \multicolumn{2}{c}{Tail classes} \\
\cmidrule{2-5}          & M-F1  & ACC   & M-F1  & ACC \\
    \midrule
    1. DNN & 0.645 & 0.772 & 0.509 & 0.651 \\
    2. DNN+Sample Reweighting & 0.658 & 0.788 & 0.548 & 0.684 \\
    3. DNN+Sample division+FixMatch & 0.712 & 0.809 & 0.534 & 0.680 \\
    4. DNN+Sample division+HeadTailMatch & 0.738 & 0.854 & 0.614 & 0.737 \\
    5. DNN+Ensemble Reweighting & 0.694 & 0.827 & 0.574 & 0.715 \\
    6. RoMaC & 0.803 & 0.871 & 0.672 & 0.784 \\
    \bottomrule
    \end{tabular}%
  \label{tab:ab}%
\end{table}%

\textbf{Result and Analysis}. 
Table \ref{tab:ab} presents the ablation study results for \texttt{RoMaC}, with FixMatch \cite{sohn2020fixmatch} as the SOTA semi-supervised algorithm. In Setting 5 (DNN + Ensemble Reweighting), each base model employs sample reweighting. Our observation and analysis are given below. 

1) By Experiment 1 and 4, the HeadTailMatch module contributes the most to performance, improving four metrics by 0.093, 0.082, 0.105, and 0.086, respectively. Improvements are particularly pronounced for tail-class samples, as high-loss tail-class samples, often misclassified due to class imbalance, are prevalent. \texttt{RoMaC} effectively corrects high-confidence erroneous pseudo-labels, mitigating this issue.

2) The results of Experiment 4 outperform those of  Experiment 3, demonstrating HeadTailMatch’s superiority in addressing both label noise and class imbalance within a semi-supervised framework. The results of Experiment 3 are also well, indicating that the semi-supervised framework is a promising direction for countering label noise.

3) Compared to the sample reweighting baseline (Experiments 2 and 5), the Ensemble Reweighting improves the four metrics by 0.036, 0.039, 0.023, and 0.039, respectively. By leveraging model diversity for complementary effects, the ensemble mitigates overfitting to head or tail data observed in single models.

4) Overall, compared with the baseline DNN, the more substantial performance gains on tail classes—an improvement of 16.3\% in M-F1 and 13.3\% in ACC—demonstrate the effectiveness of \texttt{RoMaC} in mitigating class imbalance.

\section{DISCUSSION AND LIMITATION}
\subsection{Limitation}
% \textbf{Zero-day Attacks} \texttt{RoMaC}'s design relies on the distribution characteristics of the training dataset, yet zero-day attacks introduce novel malware variants absent from this distribution, posing challenges. Although \texttt{RoMaC} excels at handling class imbalance and can theoretically adapt to new categories by incorporating new samples and retraining, this approach depends on sample availability and timely updates. Such reliance may introduce delays, limiting its effectiveness against rapidly evolving threats. Integrating online learning or anomaly detection could enhance \texttt{RoMaC}’s ability to identify novel threats without frequent retraining.

\textbf{Time Cost and Model Performance}. The ensemble strategy in \texttt{RoMaC} increases computational complexity and time cost, especially when dealing with large-scale datasets or frequent large-scale data updates. Ablation analysis shows that the core framework of \texttt{RoMaC} remains effective even without the ensemble component. In real-time scenarios (e.g., threat detection or resource-constrained edge devices), bypassing the ensemble can improve efficiency at the cost of robustness degradation, better meeting low-latency needs. However, this may slightly reduce performance on rare family samples. Future work could explore algorithmic optimizations or lightweight ensemble techniques to better balance computational cost and performance.

% \textbf{Dataset Construction}. In constructing the dataset, we treat family labels from AVLabel as clean labels. However, approximately 19\% of samples are either unclassifiable or assigned low-confidence labels by AVClass \cite{sebastian2016avclass}. While manual labeling improves accuracy, its time-consuming nature makes it impractical for large datasets. \texttt{RoMaC}, designed to handle label noise, partially mitigates the impact of these errors. Thus, when applied to manually labeled data, \texttt{RoMaC} is expected to perform better due to more reliable annotations. However, due to resource constraints, we have not fully validated this hypothesis experimentally. Future work could integrate a small set of high-quality manual labels to further enhance \texttt{RoMaC}’s performance.

\subsection{Discussion}
% \textbf{Enhancing \texttt{RoMaC} with Dynamic Features} \texttt{RoMaC} leverages static features (e.g., API call sequences and permission requests extracted) as input, achieving excellent experimental performance. However, against advanced obfuscation techniques, static features may fail to fully capture malicious behavior. Incorporating dynamic features (e.g., system calls, network activity) or testing on dynamic features could enhance \texttt{RoMaC}’s classification of complex samples and zero-day attacks. Yet, dynamic analysis is time-consuming and vulnerable to sandbox evasion, requiring a cost-benefit tradeoff. Future work could explore small-scale integration of dynamic features to validate their impact and refine feature selection.

\textbf{The Generalizability of Key Insight}. Our study reveals that the unreliability of pseudo-labels in tail classes is not confined to the malware domain. This phenomenon also persists in CIFAR-100 \cite{krizhevsky2009learning}, where the error rate of high-confidence pseudo-labels in tail-class samples reaches 73.8\%, significantly higher than in head-class samples. This finding demonstrates that, even with highly confident model predictions, pseudo-label quality remains suboptimal in tail classes, underscoring the universality of pseudo-label quality issues under long-tailed distributions. These results indirectly support the necessity and broad applicability of \texttt{RoMaC}'s differentiated processing approach for head-class and tail-class samples.

\textbf{Necessity of Class Imbalance Learning in Malware Family Classification}. Class imbalance poses a critical challenge in malware family classification, particularly due to the importance of rare, high-risk tail-class samples (e.g., APTs or new variants) that, while scarce, can cause severe security threats if missed. Traditional downsampling fails to address their extreme sparsity and may degrade performance by discarding valuable head-class samples. \texttt{RoMaC} leverages ensemble reweighting to enhance focus on rare samples, achieving robust performance under severe imbalance. Nonetheless, the limited availability of new family samples remains a challenge. Future improvements via adaptive reweighting or few-shot learning could further strengthen \texttt{RoMaC}'s capability for comprehensive protection.

\section{RELATED WORK}
\subsection{Label Noise Learning}
% Label noise is a key challenge in deep learning. 
In realistic environments, acquiring high-quality training sets is extremely costly, and training sets often inevitably contain label noise.
% obtaining high-quality labeled data is often costly and susceptible to factors such as human error or uncertainty. 
To counter label noise, researchers have proposed various label noise learning methods. For example, NoiseModel \cite{goldberger2017training} uses identity matrix initialization with regularization to ensure sparsity. ELR \cite{liu2020early} adds a regularization term to cross-entropy loss to prevent memorization of noisy labels. LRT \cite{zheng2020error} incorporates statistical hypothesis testing and Bayesian frameworks for label correction.
 MentorMix \cite{jiang2020beyond} combines curriculum learning and data augmentation to progressively remove noise. 
 
Conventional label noise learning methods face challenges in the malware domain due to the high-dimensional and complex feature space, which limits the effectiveness of generic label noise detection and correction techniques. Bai et al. \cite{wang2022malwhiteout} propose MalWhiteout, combining Confident Learning (CL) with ensemble learning to enhance noise robustness. Xu et al. \cite{D_Xu_2021} introduce Differential Training, which computes varying loss values from differently processed datasets and applies anomaly detection to correct noisy labels. In contrast, \texttt{RoMaC} improves hybrid strategies by enhancing the reliability of pseudo-labels in semi-supervised learning, thereby mitigating the negative impact of incorrect pseudo-labels on model performance.

\subsection{Class Imbalance Learning}
Class imbalance 
% refers to datasets in which minority class samples are significantly fewer than majority class samples, 
is prevalent in practical tasks such as malware family classification. 
Researchers have proposed several strategies to mitigate class imbalance issues. \textbf{Data Resampling} includes oversampling methods like SMOTE \cite{fernandez2018smote} and ADASYN \cite{he2008adasyn}, as well as undersampling approaches such as Tomek Links \cite{batista2003balancing} and NearMiss \cite{mani2003knn}. \textbf{Cost-sensitive learning} introduces class weights during training to enhance the influence of minority classes. For example, CRT \cite{kang2019decoupling} adjusts the classifier through normalization and learnable weight scaling to rebalance decision boundaries. LDAM \cite{cao2019learning} utilizes cost-sensitive learning primarily through the Label-Distribution-Aware Margin loss. 

For malware family classification, \cite{bai2020unsuccessful} proposes a Siamese network-based method, using contrastive learning to train a feature extractor. \cite{bai2021comparative} applies focal loss to prioritize difficult-to-classify malware during training. Compared to existing work, \texttt{RoMaC} introduces a novel ensemble reweighting method that addresses the drawback of single reweighting approaches—where enhancing tail classes often comes at the cost of sacrificing head class accuracy—by balancing the focus on both head and tail classes, thereby overcoming the limitations of single reweighting coefficients.

\subsection{Label Noise Learning under Class Imbalance}
Research on the interplay between label noise and class imbalance is still in its early stages. HAR \cite{cao2020heteroskedastic} employs adaptive methods to regularize noisy and tail samples. TABASCO \cite{lu2023label} separates clean and noisy samples from multiple perspectives. CBS \cite{liu2024learning} addresses imbalanced noisy labels by proposing label-balanced selection and confidence augmentation. 

In the malware family classification domain, MORSE \cite{wu2023grim} innovatively combines sample selection with semi-supervised learning, dynamically adjusting loss contributions through sample reweighting mechanisms. \texttt{RoMaC} is designed based on the understanding of the interplay between label noise and imbalance, achieving robust results by dividing potential noisy samples into head and tail categories for targeted processing.

\section{CONCLUSION}
Real-world malware datasets are characterized by significant label noise (primarily due to code obfuscation) and severe class imbalance, posing substantial challenges for Android malware family classification. Due to the interplay between label noise and class imbalance, the study of Android malware family classification should jointly address these two issues.
% These conditions underscore the need for robust classifiers capable of learning under imperfect supervision. 
Building on the understanding of their interplay, we propose a robust Android malware family classification framework \texttt{RoMaC}. It leverages a self-training algorithm to mitigate label noise under class imbalance and employs an ensemble reweighting method to handle class imbalance in the presence of label noise. Extensive experiments demonstrate that \texttt{RoMaC} achieves the great performance under varying settings, exhibiting its advantage in robustness and generalization against the existing methods.

\bibliographystyle{IEEEtran}
\bibliography{ref}

% \bibitem{ref1}
% {\it{Mathematics Into Type}}. American Mathematical Society. [Online]. Available: https://www.ams.org/arc/styleguide/mit-2.pdf

% \bibitem{ref2}
% T. W. Chaundy, P. R. Barrett and C. Batey, {\it{The Printing of Mathematics}}. London, U.K., Oxford Univ. Press, 1954.

% \bibitem{ref3}
% F. Mittelbach and M. Goossens, {\it{The \LaTeX Companion}}, 2nd ed. Boston, MA, USA: Pearson, 2004.

% \bibitem{ref4}
% G. Gr\"atzer, {\it{More Math Into LaTeX}}, New York, NY, USA: Springer, 2007.

% \bibitem{ref5}M. Letourneau and J. W. Sharp, {\it{AMS-StyleGuide-online.pdf,}} American Mathematical Society, Providence, RI, USA, [Online]. Available: http://www.ams.org/arc/styleguide/index.html

% \bibitem{ref6}
% H. Sira-Ramirez, ``On the sliding mode control of nonlinear systems,'' \textit{Syst. Control Lett.}, vol. 19, pp. 303--312, 1992.

% \bibitem{ref7}
% A. Levant, ``Exact differentiation of signals with unbounded higher derivatives,''  in \textit{Proc. 45th IEEE Conf. Decis.
% Control}, San Diego, CA, USA, 2006, pp. 5585--5590. DOI: 10.1109/CDC.2006.377165.

% \bibitem{ref8}
% M. Fliess, C. Join, and H. Sira-Ramirez, ``Non-linear estimation is easy,'' \textit{Int. J. Model., Ident. Control}, vol. 4, no. 1, pp. 12--27, 2008.

% \bibitem{ref9}
% R. Ortega, A. Astolfi, G. Bastin, and H. Rodriguez, ``Stabilization of food-chain systems using a port-controlled Hamiltonian description,'' in \textit{Proc. Amer. Control Conf.}, Chicago, IL, USA,
% 2000, pp. 2245--2249.

% \end{thebibliography}
% \input{appendix}

\end{document}